\documentclass[prd,onecolumn,nofootinbib,showpacs,showkeys]{revtex4}%
\usepackage{epsfig,graphicx,bm}
\usepackage[latin2]{inputenc}
\usepackage{color}
\usepackage{t1enc}
\usepackage{amssymb}
\usepackage{amsmath}
\usepackage{graphicx}
\usepackage{bm}
\usepackage{amsfonts}
\usepackage{array,multirow}
\begin{document}
\title{Excited Scalar and Pseudoscalar Mesons in the Extended Linear Sigma Model}
\author{$\text{Denis Parganlija}^{1}$}
\email{denisp@hep.itp.tuwien.ac.at}
\author{$\text{Francesco Giacosa}^{2,3}$}
\email{fgiacosa@ujk.edu.pl}
\affiliation{$^{1}$Institut f\"{u}r Theoretische Physik, 
Technische Universit\"{a}t Wien, Wiedner Hauptstr.\ 8-10, 1040 Vienna, Austria}
\affiliation{$^{2}$Institute of Physics, Jan Kochanowski University, 
ul.\ Swietokrzyska 15, 25-406 Kielce, Poland}
\affiliation{$^{3}$Institut f\"{u}r Theoretische Physik, Johann Wolfgang 
Goethe-Universit\"{a}t, Max-von-Laue-Str.\ 1, 60438 Frankfurt am Main, Germany}

\begin{abstract}
We present an in-depth study of masses and decays of
excited scalar and pseudoscalar $\bar{q}q$ states in the
Extended Linear Sigma Model (eLSM). The model also contains ground-state scalar,
pseudoscalar, vector and axial-vector mesons. The main objective is
to study the consequences of the hypothesis that the $f_0(1790)$
resonance, observed a decade ago by the BES Collaboration
and recently by LHCb, represents an excited scalar quarkonium.
In addition we also analyse the possibility that the new $a_0(1950)$
resonance, observed recently by BABAR, may also be an excited scalar state.
Both hypotheses receive justification in our approach
although there appears to be some tension between the simultaneous
interpretation of $f_0(1790)$/$a_0(1950)$ and pseudoscalar mesons
$\eta(1295)$, $\pi(1300)$, $\eta(1440)$ and $K(1460)$ as excited
$\bar{q}q$ states.
\end{abstract}

\maketitle

\allowdisplaybreaks

\section{Introduction}

One of the most important features of strong interaction is the existence of
the hadron spectrum. It emerges from confinement of
quarks and gluons -- degrees of freedom of the underlying theory,
Quantum Chromodynamics (QCD) -- in regions of sufficiently low energy 
where the QCD coupling is known to be large
\cite{AF1,AF2,AF3,AF4}. Although the exact mechanism of hadron 
formation in non-perturbartive QCD is
not yet fully understood, an experimental fact is a very abundant spectrum of
states possessing various quantum numbers, such as for example isospin $I$,
total spin $J$, parity $P$ and charge conjugation $C$.\newline\newline 
This is in particular the case for the spectrum of mesons (hadrons with integer
spin) that can be found in the listings of PDG -- the Particle Data Group
\cite{PDG}. 
In the scalar channel ($J^{P}=0^{+}$), the following states
are listed in the energy region up to approximately 2 GeV:
\begin{align*}
& f_{0}(500)/\sigma\text{, }K_{0}^{\star}(800)/\kappa\text{, }a_{0}%
(980)\text{, }f_{0}(980)\text{, }f_{0}(1370)\text{, }K_{0}^{\star
}(1430)\text{, }a_{0}(1450)\text{, }f_{0}(1500)\text{, }f_{0}(1710)\text{,
}\\
& K_{0}^{\star}(1950)\text{, }a_{0}(1950)\text{, }f_{0}(2020)\text{, }%
f_{0}(2100)\; .
\end{align*}
The pseudoscalar channel ($J^{P}=0^{-}$) is similarly well populated:
\[
\pi\text{, }K\text{, }\eta\text{, }\eta^{\prime}(958)\text{, }\eta
(1295)\text{, }\pi(1300)\text{, }\eta(1405)\text{, }K(1460)\text{, }%
\eta(1475)\text{, }\eta(1760)\text{, }\pi(1800)\text{, }K(1830) \;.
\]
A natural expectation founded in the Quark Model
(see Refs.\ \cite{GM,Zweig}; for a modern and modified version see
for example Refs.\ \cite{Rupp,Zacchi:2016tjw}) is that the mentioned states
can effectively be described in terms of constituent quarks and antiquarks
-- ground-state $\bar{q}q$ resonances. In this context, we define ground 
states as those with the lowest mass for a given set of quantum numbers $I$, $J$, $P$ and $C$.
Such description is particularly successful for the
lightest pseudoscalar states $\pi$, $K$ and $\eta$.
\\
\\
However, this cannot be the full picture as the spectra contain
more states than could be described in terms of the ground-state
$\bar{q}q$ structure. A further natural expectation is then that the spectra
may additionally contain first (radial) excitations of $\bar{q}q$ states,
i.e., those with the same quantum numbers but with higher masses.
(In the spectroscopic notation, the excited scalar and pseudoscalar states 
correspond respectively to the $2\, ^{3}P_0$ and $2\, ^{1}S_0$ configurations.)
Of course, the possibility to study such states depends crucially
on the identification of the ground states themselves; in the case of
the scalar mesons, this is not as clear as for the pseudoscalars.
Various hypotheses have been suggested for the scalar-meson structure,
including meson-meson molecules, 
$\bar{q}\bar{q}qq$ states and glueballs, bound states of gluons --
see, e.g., Refs.\
\cite{Basdevant:1972uu,Estabrooks:1978de,Ohta1,Ohta2,Ohta3,Close1988,vanBeveren:1986ea,Zou:1994ea,
Kaminski:1993zb,Achasov:1994iu,Tornqvist:1995,Muenz:1996,Dobado:1996ps,Elias:1998bq,
Black:1998wt,Minkowski:1998mf,Oller:1998zr,Kaminski:1998ns,Ishida:1999qk,Surovtsev:2000ev,
Black:2000qq,Teshima2002,scalars-above1GeVqq-below1GeVq2q2,Anisovich-KMatrix,
Pelaez2003,Bugg:2003kj,Scadron:2003yg,Napsuciale2004,Pelaez:2004xp,Lattice3,
Lattice5,Fariborz:2007ai,Albaladejo:2008qa,Fariborz2009,Mennessier:2010xg,
Branz:2010gd,GarciaMartin:2011jx,Mukherjee:2012xn,Fariborz:2015era,
Eichmann:2015cra,Pelaez:2015qba,Fariborz:2015dou,Kirchbach:2016scz,
Amsler1995,LatticeGlueball1,Lee1999,Close2001,
Amsler2004,LatticeGlueball2,Close2005,Giacosa2005,LatticeGlueball3,Forkel2007,
Mathieu2008,LatticeGlueball4,LatticeGlueball5,Brunner:2015oqa,Brunner:2015yha,Capossoli1,
Brunner:2015oga,Capossoli2,Capossoli3,Capossoli4,Lattice1,Lattice2,Lattice4,Lattice6}.
Results of these studies are at times conflicting but the general conclusion is
nonetheless that the scalar $\bar{q}q$ ground states
(as well as the glueball and the low-energy four-quark states) are well-defined 
and positioned in the spectrum of particles up to and including the $f_0(1710)$ resonance.
\\
\\
The main objective of this work is then to ascertain which properties the excited
scalar and pseudoscalar $\bar{q}q$ states possess and whether they can be identified 
in the physical spectrum.
\\
\\
Our study of the excited mesons is based on the Linear Sigma
Model \cite{GellMann:1960np,geffen,ko,urban}. This is an effective approach
to low-energy QCD -- its degrees of freedom are not quarks and gluons
of the underlying theory but rather meson fields with various values of
$I$, $J$, $P$ and $C$. \\
There are several advantages that the model has to offer. Firstly, it implements the symmetries of QCD as well as 
their breaking (see Sec.\ \ref{A0} for details). Secondly, it contains degrees of freedom with quantum numbers equal to those observed experimentally and in theoretical first-principles spectra (such as those of lattice QCD). This combination of symmetry-governed dynamics and states with correct quantum numbers justifies in our view the expectation that important aspects of the strong interaction are captured by the proposed model. Note that the model employed in this article is wide-ranging in that it contains the ground-state scalar, pseudoscalar, vector and axial-vector $\bar{q}q$ states in three flavours
($u$, $d$, $s$), the scalar dilaton (glueball) and the first excitations in the 
three-flavour scalar and pseudoscalar channels. 
Considering isospin multiplets as single degrees of freedom, there are
16 $\bar{q}q$ ground states and 8 $\bar{q}q$ excited states plus the scalar
glueball in the model.
For this reason, it can be denoted as the "Extended Linear Sigma Model" (eLSM). A further advantage of eLSM is that the inclusion of degrees of freedom with a certain structure (such as $\bar{q}q$ states here) allows us to test the compatibility of experimentally known resonances with such structure. This is of immediate relevance for experimental hadron searches such as those planned at PANDA@FAIR \cite{PANDA}. \\
With regard to vacuum states,
the model has been used in studies of two-flavour $\bar{q}q$
mesons \cite{Article2010}, glueballs \cite{Stani1,Eshraim:2012jv,Eshraim:2016mds,Giacosa:2016hrm,Koenigstein:2016tjw}, $K_1$ and other spin-1 mesons \cite{Divotgey:2013jba,Habersetzer:2015eha} and
baryons \cite{Gallas:2009qp}. It is, however, also suitable for studies of the QCD phase diagram \cite{Lenaghan,RS,Wolf}. In this article, we will build upon the results obtained in Refs.\ \cite{Article2013,Stani2} where ground-state $\bar{q}q$ resonances and the glueball were considered in vacuum. Comparing experimental masses and decay widths with the theoretical predictions for excited states, we will draw conclusions on structure of the observed states; we will also predict more than 35 decays for various scalar and pseudoscalar resonances (see Sec.\ \ref{Decays2}).
\\
\\
Irrespective of the above advantages, we must note that the model used in this article also has drawbacks. There are two that appear to be of particular importance. \\
Firstly, some of the states that might be of relevance in the region of interest are absent. The most important example is the scalar glueball whose mass is comparable \cite{LatticeGlueball1,LatticeGlueball2,LatticeGlueball3,LatticeGlueball4,LatticeGlueball5} to that of the excited $\bar{q}q$ states discussed here. The implementation of the scalar glueball is actually straightforward in our approach (see Sec.\ \ref{A0}) but the amount of its mixing with excited states is as yet unestablished, mainly due to the unfortunate lack of experimental data (discussed in Sec.\ \ref{LagrangianS}).\\
Secondly, our calculations of decay widths are performed at tree level. Consequently, unitarity corrections are not included. A systematic way to implement them is to consider mesonic loops and determine their influence on the pole positions of resonances. Substantial shift of the pole position may then improve (or spoil) the comparison to the experimental data. However, the results of Ref.\ \cite{Pagliara} suggest that unitarity corrections are small for resonances whose ratio of decay width to mass is small as well. Since such resonances are present in this article (see Sec.\ \ref{Decays5}), the corrections will not be considered here.
\\
\\
Excited mesons were a subject of interest already several decades ago 
\cite{Freund:1968nf,ChenTsai:1974cn}; to date, they have been considered in 
a wide range of approaches including QCD models / chiral Lagrangians
\cite{Ino:1984pi,Geiger:1993fb,Fedorov:2003zy,Vijande:2005jd,Gutsche:2008qq,Rupp:2016uee}, 
Lattice QCD \cite{Lacock:1996vy,Burch:2006dg,Dudek:2009qf,Dudek:2010wm,Engel:2011aa,
Dudek:2013yja}, Bethe-Salpeter equation 
\cite{Holl:2004fr,Holl:2004un,Holl:2005vu,Li:2016dzv}, NJL Model and its extensions 
\cite{Volkov:1997dd,Volkov:1998fa,Volkov:1999iq,Volkov:1999yi,Volkov:1999xf,
Volkov:2000ry,Volkov:2001ct,Volkov:2002iz,Arbuzov:2010vq,Vishneva:2013mga,
Vishneva:2014efa}, 
light-cone models \cite{Arndt:1999wx}, QCD string approaches
\cite{Badalian:2002rc} and QCD domain walls \cite{Nedelko:2016gdk}. 
Chiral symmetry has also been suggested to become effectively
restored in excited mesons \cite{Wagenbrunn:2006cs,Wagenbrunn:2007ie}
rendering their understanding even more important. 
A study analogous to ours (including both scalar and pseudoscalar excitations and their various decay channels) was performed in extensions of the NJL model \cite{Volkov:1999iq,Volkov:1999yi,Volkov:1999xf,Volkov:2001ct,Volkov:2002iz}. The conclusion was that $f_0(1370)$, $f_0(1710)$ and $a_0(1450)$ are the first radial excitations of $f_0(500)$, $f_0(980)$ and $a_0(980)$. However, this is at the expense of having very large decay widths for $f_0(1370)$, $f_0(1500)$ and $f_0(1710)$; in our case the decay widths for $f_0$ states above 1 GeV correspond to experimental data but the resonances are identified as quarkonium ground states \cite{Article2013}.
\\
\\
The outline of the article is as follows. The general structure and results obtained
so far regarding ground-state $\bar{q}q$ resonances are briefly reviewed in Sections
\ref{A} and \ref{B}. Building upon that basis, we present the Lagrangian
for the excited states and discuss the relevant experimental data in
Sec.\ \ref{C}. Two hypotheses are tested in Sec.\ \ref{Results}: 
whether the $f_0(1790)$ and $a_0(1950)$ resonances can represent
excited $\bar{q}q$ states; the first one is not (yet) listed by the PDG 
but has been observed
by the BES II and LHCb Collaborations \cite{BESII,LHCb} and is discussed
in Sec.\ \ref{LagrangianS}. We also discuss to what extent it is possible to
interpret the pseudoscalar mesons $\eta(1295)$, $\pi(1300)$, $\eta(1440)$
and $K(1460)$ as excited states. 
Conclusions are presented in Sec.\ \ref{Conclusion} and all interaction
Lagrangians used in the model can be found in Appendix \ref{Appendix}.
Our units are $\hbar = c = 1$; the metric tensor is $g_{\mu \nu} =$
diag$(+,-,-,-)$.

\section{The Model} \label{A0}

\subsection{General Remarks} \label{A}

A viable effective approach to phenomena of non-perturbative strong
interaction must implement the symmetries present in the underlying theory,
QCD. The theory itself is rich in symmetries: colour symmetry $SU(3)_{c}$
(local); chiral $U(N_{f})_{L}\times U(N_{f})_{R}$ symmetry ($L$ and $R$ denote
the 'left' and 'right' components and $N_{f}$ the number of quark flavours;
global, broken in vacuum spontaneously by the non-vanishing chiral condensate
$\langle\bar{q}q\rangle$ \cite{SSB1,SSB2}, at the quantum level via the axial
$U(1)_{A}$\ anomaly \cite{Hooft} and explicitly by the non-vanishing quark
masses); dilatation symmetry (broken at the quantum level
\cite{Marciano,Crewther} but valid classically in QCD without quarks); $CPT$
symmetry (discrete; valid individually for charge conjugation $C$, parity
transformation $P$ and time reversal $T$); $Z_{3}$ symmetry (discrete;
pertaining to the centre elements of a special unitary matrix of dimension
$N_{f}\times N_{f}$; non-trivial only at non-zero temperatures
\cite{Roberge,Z31,Z32,Z33,Z34,Kogan:1998zc}) -- all of course in addition to the 
Poincar\'{e} symmetry.\newline
Terms entering the Lagrangian of an effective approach to QCD should as a
matter of principle be compatible with all symmetries listed above. Our subject is
QCD in vacuum. In this context, we note that the 
colour symmetry
is automatically fulfilled since we will be working with colour-neutral degrees of
freedom; the structure and number of terms entering
the Lagrangian are then restricted by the chiral, CPT and dilatation
symmetries. \newline
\\
\\
The eLSM Lagrangian has the following general structure:
\begin{equation}
\mathcal{L}=\mathcal{L}_{dil.}+\mathcal{L}_{0}+\mathcal{L}_{E} \label{L}%
\end{equation}
and in Sections \ref{B} and \ref{C} we discuss the structure of the Lagrangians
contributing to $\mathcal{L}$ as well as their matter content.

\subsection{Ground-State Quarkonia and Dilaton: Lagrangian and the Matter
Content} \label{B}

This section contains a brief overview of the results obtained so far in the
Extended Linear Sigma Model that contains $N_{f}=3$ scalar, pseudoscalar,
vector and axial-vector quarkonia and the scalar glueball. The discussion is
included for convenience of the reader and in order to set the basis for the
incorporation of the excited quarkonia (Sec.\ \ref{C}). All details can be found in
Refs.\ \cite{Article2013,Stani2}. \newline
\\
In Eq.\ (\ref{L}), $\mathcal{L}_{dil}$ implements, at the composite level, the
dilatation symmetry of QCD and its breaking
\cite{Rosenzweig:1979ay,Rosenzweig:1981cu,
Rosenzweig:1982cb,Migdal:1982jp,Gomm:1984zq,Gomm:1985ut}:%
\begin{equation}
\mathcal{L}_{dil.}=\frac{1}{2}(\partial_{\mu}G)^{2}-\frac{1}{4}\frac{m_{G}%
^{2}}{\Lambda^{2}}\left(  G^{4}\ln\frac{G^{2}}{\Lambda^{2}}-\frac{G^{4}}%
{4}\right)  \label{ldil}%
\end{equation}
where $G$ represents the dilaton field and $\Lambda$ is the scale that
explicitly breaks the dilatation symmetry. Considering fluctuations around the
potential minimum $G_{0}\equiv\Lambda$ leads to the emergence of a particle
with $J^{PC}=0^{++}$ -- the scalar glueball \cite{Stani2,Stani1}. \newline
\\
Terms that (\textit{i}) are compatible in their structure with the chiral,
dilatation and CPT symmetries of QCD and (\textit{ii}) contain ground-state
scalar, pseudoscalar, vector and axial-vector quarkonia with $N_{f}=3$ and the
dilaton are collected in the $\mathcal{L}_{0}$ contribution to Eq.\ (\ref{L}),
as in Refs.\ \cite{Article2010,Stani2,Article2013}:
\begin{align}
\mathcal{L}_{0}  &
=\mathop{\mathrm{Tr}}[(D_{\mu}\Phi)^{\dagger}(D_{\mu}\Phi)]-m_{0}^{2}\left(
\frac{G}{G_{0}}\right)  ^{2}\mathop{\mathrm{Tr}}(\Phi^{\dagger}\Phi
)-\lambda_{1}[\mathop{\mathrm{Tr}}(\Phi^{\dagger}\Phi)]^{2}-\lambda
_{2}\mathop{\mathrm{Tr}}(\Phi^{\dagger}\Phi)^{2}{\nonumber}\nonumber\\
&  -\frac{1}{4}\mathop{\mathrm{Tr}}(L_{\mu\nu}^{2}+R_{\mu\nu}^{2}%
)+\mathop{\mathrm{Tr}}\left[  \left(  \left(  \frac{G}{G_{0}}\right)
^{2}\frac{m_{1}^{2}}{2}+\Delta\right)  (L_{\mu}^{2}+R_{\mu}^{2})\right]
+\mathop{\mathrm{Tr}}[H(\Phi+\Phi^{\dagger})]+\mathop{\mathrm{Tr}}(\Phi
^{\dagger}\Phi E_{0}+\Phi\Phi^{\dagger}E_{0}){\nonumber}\nonumber\\
&  +c_{1}(\det\Phi-\det\Phi^{\dagger})^{2}+i\frac{g_{2}}{2}%
(\mathop{\mathrm{Tr}}\{L_{\mu\nu}[L^{\mu},L^{\nu}%
]\}+\mathop{\mathrm{Tr}}\{R_{\mu\nu}[R^{\mu},R^{\nu}]\}){\nonumber}\nonumber\\
&  +\frac{h_{1}}{2}\mathop{\mathrm{Tr}}(\Phi^{\dagger}\Phi
)\mathop{\mathrm{Tr}}(L_{\mu}^{2}+R_{\mu}^{2})+h_{2}%
\mathop{\mathrm{Tr}}[\vert L_{\mu}\Phi \vert ^{2}+\vert \Phi R_{\mu} \vert ^{2}]+2h_{3}%
\mathop{\mathrm{Tr}}(L_{\mu}\Phi R^{\mu}\Phi^{\dagger}){\nonumber}\nonumber\\
&  +g_{3}[\mathop{\mathrm{Tr}}(L_{\mu}L_{\nu}L^{\mu}L^{\nu}%
)+\mathop{\mathrm{Tr}}(R_{\mu}R_{\nu}R^{\mu}R^{\nu})]+g_{4}%
[\mathop{\mathrm{Tr}}\left(  L_{\mu}L^{\mu}L_{\nu}L^{\nu}\right)
+\mathop{\mathrm{Tr}}\left(  R_{\mu}R^{\mu}R_{\nu}R^{\nu}\right)
]{\nonumber}\nonumber\\
&  +g_{5}\mathop{\mathrm{Tr}}\left(  L_{\mu}L^{\mu}\right)
\,\mathop{\mathrm{Tr}}\left(  R_{\nu}R^{\nu}\right)  +g_{6}%
[\mathop{\mathrm{Tr}}(L_{\mu}L^{\mu})\,\mathop{\mathrm{Tr}}(L_{\nu}L^{\nu
})+\mathop{\mathrm{Tr}}(R_{\mu}R^{\mu})\,\mathop{\mathrm{Tr}}(R_{\nu}R^{\nu
})]\text{ .} \label{Lagrangian}%
\end{align}
$\,$
\\
In Eq.\ (\ref{Lagrangian}), the matrices $\Phi$, $L^{\mu}$, and $R^{\mu}$
represent the scalar and vector nonets:
\begin{align}
\Phi &  =\sum_{i=0}^{8}(S_{i}+iP_{i})T_{i}=\frac{1}{\sqrt{2}}\left(
\begin{array}
[c]{ccc}%
\frac{(\sigma_{N}+a_{0}^{0})+i(\eta_{N}+\pi^{0})}{\sqrt{2}} & a_{0}^{+}%
+i\pi^{+} & K_{0}^{\star+}+iK^{+}\\
a_{0}^{-}+i\pi^{-} & \frac{(\sigma_{N}-a_{0}^{0})+i(\eta_{N}-\pi^{0})}%
{\sqrt{2}} & K_{0}^{\star0}+iK^{0}\\
K_{0}^{\star-}+iK^{-} & {\bar{K}_{0}^{\star0}}+i{\bar{K}^{0}} & \sigma
_{S}+i\eta_{S}%
\end{array}
\right)  \text{ ,}\label{eq:matrix_field_Phi}\\
L^{\mu}  &  =\sum_{i=0}^{8}(V_{i}^{\mu}+A_{i}^{\mu})T_{i}=\frac{1}{\sqrt{2}%
}\left(
\begin{array}
[c]{ccc}%
\frac{\omega_{N}+\rho^{0}}{\sqrt{2}}+\frac{f_{1N}+a_{1}^{0}}{\sqrt{2}} &
\rho^{+}+a_{1}^{+} & K^{\star+}+K_{1}^{+}\\
\rho^{-}+a_{1}^{-} & \frac{\omega_{N}-\rho^{0}}{\sqrt{2}}+\frac{f_{1N}%
-a_{1}^{0}}{\sqrt{2}} & K^{\star0}+K_{1}^{0}\\
K^{\star-}+K_{1}^{-} & {\bar{K}}^{\star0}+{\bar{K}}_{1}^{0} & \omega
_{S}+f_{1S}%
\end{array}
\right)  ^{\mu}\text{,}\label{eq:matrix_field_L}\\
R^{\mu}  &  =\sum_{i=0}^{8}(V_{i}^{\mu}-A_{i}^{\mu})T_{i}=\frac{1}{\sqrt{2}%
}\left(
\begin{array}
[c]{ccc}%
\frac{\omega_{N}+\rho^{0}}{\sqrt{2}}-\frac{f_{1N}+a_{1}^{0}}{\sqrt{2}} &
\rho^{+}-a_{1}^{+} & K^{\star+}-K_{1}^{+}\\
\rho^{-}-a_{1}^{-} & \frac{\omega_{N}-\rho^{0}}{\sqrt{2}}-\frac{f_{1N}%
-a_{1}^{0}}{\sqrt{2}} & K^{\star0}-K_{1}^{0}\\
K^{\star-}-K_{1}^{-} & {\bar{K}}^{\star0}-{\bar{K}}_{1}^{0} & \omega
_{S}-f_{1S}%
\end{array}
\right)  ^{\mu}\text{,} \label{eq:matrix_field_R}%
\end{align}
where $T_{i}\,(i=0,\ldots,8)$ denote the generators of $U(3)$, while $S_{i}$
represents the scalar, $P_{i}$ the pseudoscalar, $V_{i}^{\mu}$ the vector,
$A_{i}^{\mu}$ the axial-vector meson fields.
[Note that we are using the non-strange--strange basis
defined as $\varphi_{N}=\frac{1}{\sqrt{3}}\left(  \sqrt{2}\;\varphi
_{0}+\varphi_{8}\right)  $ and $\varphi_{S}=\frac{1}{\sqrt{3}}\left(
\varphi_{0}-\sqrt{2}\;\varphi_{8}\right)  $ with $\varphi\in(S_{i},P_{i}%
,V_{i}^{\mu},A_{i}^{\mu})$.]
\\
\\
Furthermore,
\begin{equation}
D^{\mu}\Phi\equiv\partial^{\mu}\Phi-ig_{1}(L^{\mu}\Phi-\Phi R^{\mu})
\label{Dmu}%
\end{equation}
is the derivative of $\Phi$ transforming covariantly with regard to the
$U(3)_{L}\times U(3)_{R}$ symmetry group; the left-handed and right-handed
field strength tensors $L^{\mu\nu}$ and $R^{\mu\nu}$ are respectively defined as%
\begin{align}
L^{\mu\nu}  &  \equiv\partial^{\mu}L^{\nu}
-
\partial^{\nu}L^{\mu}
\;\text{,}%
\label{Lmunu}\\
R^{\mu\nu}  &  \equiv\partial^{\mu}R^{\nu}
-
\partial^{\nu}R^{\mu}
\;\text{.}
\label{Rmunu}%
\end{align}
\\
The following symmetry-breaking mechanism is implemented:

\begin{itemize}
\item The spontaneous breaking of the $U(3)\times U(3)$ chiral symmetry
requires setting $m_{0}^{2}<0$.

\item The explicit breaking of the $U(3)\times U(3)$ chiral as well as
dilatation symmetries is implemented by terms describing non-vanishing quark
masses: $H=$ diag$\{h_{N},h_{N},h_{S}\}$, $\Delta=$ diag$\{0,0,\delta_{S}\}$
and $E_{0}=$ diag$\{0,0,\epsilon_{S}\}$.

\item The $U(1)_{A}$ (chiral) anomaly is implemented by the determinant term
$c_{1}(\det\Phi-\det\Phi^{\dagger})^{2}$ \cite{WV1,WV2}.
\end{itemize}
$\,$\\
We also note the following important points:

\begin{itemize}
\item All states present in the Lagrangian (\ref{Lagrangian}), except for the 
dilaton, possess the $\bar{q}q$ structure \cite{Article2010,PhD}. The argument 
is essentially
based on the large-$N_{c}$ behaviour of the model parameters and on the model
construction in terms of the underlying (constituent) quark fields.
The ground-state Lagrangian (\ref{Lagrangian}) contains a pseudoscalar field assigned to the pion since it emerges from spontaneous breaking of the (chiral) $U(3) \times U(3)$ symmetry. Furthermore, the vector meson decaying into $2\pi$ is identified with the rho since the latter is experimentally known to decay into pions with a branching ratio of slightly less than 1. Pion and rho can be safely assumed to represent (very predominant) $\bar{q}q$ states and hence the large-$N_c$ behaviour of their mass terms has to be $N_c^0$. Additionally, the rho-pion vertex has to scale as $N_c^{-1/2}$ since the states are quarkonia. Then, as shown in Ref.\ \cite{Article2010}, this is sufficient to determine the large-$N_c$ behaviour of all ground-state model parameters and of the non-strange and strange quark condensates. As a consequence, the masses of all other ground states scale as $N_c^0$ and their decay widths scale as $1/N_c$. For this reason, we identify these degrees of freedom with $\bar{q}q$ states.\\
A further reason is that all states entering the matrix $\Phi$ in Eq.\ (\ref{eq:matrix_field_Phi}) can be decomposed in terms of (constituent) quark currents whose behaviour under chiral transformation is such that all terms in the Lagrangian (except for symmetry-breaking or anomalous ones) are chirally symmetric \cite{PhD}.\\
Note that our excited-state Lagrangian (\ref{LagrangianE}) will have exactly the same structure as the ground-state one.
Considering the above discussion, we conclude that its degrees of freedom also have the $\bar{q}q$ structure.

\item The number of terms entering Eq.\ (\ref{Lagrangian}) is finite under the
requirements that (\textit{i}) all terms are dilatationally invariant and
hence have mass dimension equal to four, except possibly for those that are
explicitly symmetry-breaking or anomalous, and (\textit{ii}) no term leads to
singularities in the potential in the limit $G\rightarrow0$ \cite{dynrec}.

\item Notwithstanding the above point, the glueball will not be a subject of
this work -- hence $G \equiv G_{0}$ is set throughout this article. With
regard to the ground-state mesons, we will be relying on
Ref.\ \cite{Article2013} since it contains the latest results from the model
without the glueball. (For the model version with three-flavour $\bar{q}q$ states 
as well as the scalar glueball, see Ref.\ \cite{Stani2}.)

\item There are two scalar isospin-0 fields in the Lagrangian
(\ref{Lagrangian}): $\sigma_{N}\equiv\bar{n}n$ ($n$: $u$ and $d$ quarks,
assumed to be degenerate) and $\sigma_{S}\equiv\bar{s}s$. Spontaneous breaking
of the chiral symmetry implies the existence of their respective vacuum
expectation values $\phi_{N}$ and $\phi_{S}$. As described in
Ref.\ \cite{Article2013}, shifting of $\sigma_{N,S}$ by $\phi_{N,S}$ leads to
the mixing of spin-1 and spin-0 fields. These mixing terms are removed by
suitable shifts of the spin-1 fields that have the following general
structure:%
\begin{equation}
V^{\mu}\rightarrow V^{\mu}+Z_{S}w_{V}\partial^{\mu}S\;\text{,} \label{Vmu}%
\end{equation}

where $V^{\mu}$ and $S$ respectively denote the spin-1 and spin-0 fields. The
new constants $Z_{S}$ and $w_{V}$ are field-dependent and read
\cite{Article2013}:
\begin{align}
w_{f_{1N}}  &  =w_{a_{1}}=\frac{g_{1}\phi_{N}}{m_{a_{1}}^{2}}\text{ ,
\ }w_{f_{1S}}=\frac{\sqrt{2}g_{1}\phi_{S}}{m_{f_{1S}}^{2}}\text{ ,
\ }w_{K^{\star}}=\frac{ig_{1}(\phi_{N}-\sqrt{2}\phi_{S})}{2m_{K^{\star}}^{2}%
}\text{ , \ }w_{K_{1}}=\frac{g_{1}(\phi_{N}+\sqrt{2}\phi_{S})}{2m_{K_{1}}^{2}%
}\text{ ,}\\
Z_{\pi}  &  =Z_{\eta_{N}}=\frac{m_{a_{1}}}{\sqrt{m_{a_{1}}^{2}-g_{1}^{2}%
\phi_{N}^{2}}}\text{ , \ }Z_{K}=\frac{2m_{K_{1}}}{\sqrt{4m_{K_{1}}^{2}%
-g_{1}^{2}(\phi_{N}+\sqrt{2}\phi_{S})^{2}}}\text{ ,}\\
Z_{\eta_{S}}  &  =\frac{m_{f_{1S}}}{\sqrt{m_{f_{1S}}^{2}-2g_{1}^{2}\phi
_{S}^{2}}}\text{ , \ }Z_{K_{0}^{\star}}=\frac{2m_{K^{\star}}}{\sqrt
{4m_{K^{\star}}^{2}-g_{1}^{2}(\phi_{N}-\sqrt{2}\phi_{S})^{2}}}\text{ .}
\label{ZetaS}%
\end{align}
As demonstrated in Ref.\ \cite{Article2013}, $\phi_{N}$ and $\phi_{S}$ are
functions of $Z_{\pi}$ and $Z_{K}$ as follows:%
\begin{align}
\phi_{N}  &  =Z_{\pi}f_{\pi}\text{ , \ }\label{PhiN}\\
\phi_{S}  &  =\sqrt{2}Z_{K}f_{K}-\phi_{N}/\sqrt{2}\; \label{PhiS}%
\end{align}
where $f_{\pi}$ and $f_{K}$ respectively denote the pion and kaon decay constants.
\end{itemize}
$\,$
\\
The ground-state mass terms can be obtained from Lagrangian (\ref{Lagrangian}%
); their explicit form can be found in Ref.\ \cite{Article2013} where a
comprehensive fit of the experimentally known meson masses was performed. Fit
results that will be used in this article are collected in Table
\ref{Comparison1}. The following is of importance here:

\begin{itemize}
\item Table \ref{Comparison1} contains no statement on masses and assignment
of the isoscalar states $\sigma_{N}$ and $\sigma_{S}$. The reason is that their
identification in the meson spectrum is unclear due to both theoretical and
experimental uncertainties \cite{Berlin,Hersonissos}. In
Ref.\ \cite{Article2013}, the preferred assignment of $\sigma_{N}$ was to
$f_{0}(1370)$, not least due to the best-fit result $m_{\sigma_{N}} = 1363$
MeV. The resonance $\sigma_{S}$ was assigned to $f_{0}(1710)$. Note that a
subsequent analysis in Ref.\ \cite{Stani2}, that included the scalar glueball,
found the assignment of $\sigma_{S}$ to
$f_{0}(1500)$ more preferable; $f_0(1710)$ was found to be compatible with
the glueball. These issues will be of secondary importance
here since no mixing between excited and ground states
will be considered. [We also note that decays of the excited states into $f_0(1500)$ and $f_0(1710)$
would be kinematically forbidden. Excited-state masses are discussed in Sec.\ \ref{Results}.]

\item Table \ref{Comparison1} also contains no statement on the axial-vector
kaon $K_{1}$. Ref.\ \cite{Article2013} obtained $m_{K_{1}} = 1282$ MeV as the
best-fit result. One needs to note, however, that PDG listings \cite{PDG}
contain two states to which our $K_{1}$ resonance could be assigned:
$K_{1}(1270)$ and $K_{1}(1400)$. Both have a significant mutual overlap
\cite{Carnegie:1977uz,Rosner:1985dx,Isgur:1989vq,Blundell:1995au,Close1++,Burakovsky,Li:2000dy,Asner:2000nx,Carvalho,Cheng1,
Barnes:2002mu,Li,Vijande:2004he,Li:2005eq,Li:2006we,Hatanaka:2008xj,Cheng:2009ms,Yang,Cheng2}%
; analysis from the Linear Sigma Model suggests that our $K_{1}$ state has a
larger overlap with $K_{1}(1400)$ \cite{Divotgey:2013jba}. Nonetheless, we
will use $m_{K_{1}} = 1282$ MeV for decays of excited states involving $K_{1}$
-- this makes no significant difference to our results since the decays with
$K_{1}$ final states are phase-space suppressed for the mass range of excited
mesons.

\item The states $\eta$ and $\eta^{\prime}$ arise from mixing of $\eta_{N}$
and $\eta_{S}$ in Lagrangian (\ref{Lagrangian}). The mixing angle is
$\theta_{\eta}=-44.6^{\circ}$ \cite{Article2013}; see also
Refs.\ \cite{Feldmann,Feldmann:1998sh,Feldmann:1999uf,
Feldmann:2002kz,etaM,Kupsc:2007ce,Ambrosino,AmelinoCamelia:2010me,Chang:2012gnb}%
.
\end{itemize}

\begin{table}[th]
\centering
\begin{tabular}
[c]{|c|c|c|c|}\hline
Observable & Model ground state assigned to & Fit [MeV] & Experiment
[MeV]\\\hline
$m_{\pi}$ & pion & $141.0 \pm5.8$ & $137.3 \pm6.9$\\\hline
$m_{K}$ & kaon & $485.6 \pm3.0$ & $495.6 \pm24.8$\\\hline
$m_{\eta}$ & $\eta$ & $509.4 \pm3.0$ & $547.9 \pm27.4$\\\hline
$m_{\eta^{\prime}}$ & $\eta^{\prime}(958)$ & $962.5 \pm5.6$ & $957.8 \pm
47.9$\\\hline
$m_{\rho} \equiv m_{\omega_{N}}$ & $\rho(770) $ & $783.1 \pm7.0$ & $775.5
\pm38.8$\\\hline
$m_{K^{\star}}$ & $K^{\star}(892)$ & $885.1 \pm6.3$ & $893.8 \pm44.7$\\\hline
$m_{\phi}$ & $\phi(1020)$ & $975.1 \pm6.4$ & $1019.5 \pm51.0$\\\hline
$m_{a_{1}} \equiv m_{f_{1N}}$ & $a_{1}(1260)$ & $1186 \pm6$ & $1230 \pm
62$\\\hline
$m_{f_{1S}}$ & $f_{1}(1420)$ & $1372.5 \pm5.3$ & $1426.4 \pm71.3$\\\hline
$m_{a_{0}}$ & $a_{0}(1450)$ & $1363 \pm1$ & $1474 \pm74$\\\hline
$m_{K_{0}^{\star}}$ & $K_{0}^{\star}(1430)$ & $1450 \pm1$ & $1425 \pm
71$\\\hline
$f_{\pi}$ & - & $96.3 \pm0.7 $ & $92.2 \pm4.6$\\\hline
$f_{K}$ & - & $106.9 \pm0.6$ & $110.4 \pm5.5$\\\hline
\end{tabular}
\caption{Best-fit results for masses of ground-state mesons and pseudoscalar
decay constants present in Eq.\ (\ref{Lagrangian}), obtained in
Ref.\ \cite{Article2013}. The values in the third column will be used in this
article in order for us to remain model-consistent. Note that the errors in the
fourth column correspond either to the experimental values or to 
5\% of the respective central values (whichever is larger).}%
\label{Comparison1}%
\end{table}

\subsection{Excited Scalars and Pseudoscalars} \label{C}

\subsubsection{Lagrangian}

\label{LagrangianS}

With the foundations laid in the previous section, the most general Lagrangian
for the excited scalar and pseudoscalar quarkonia with terms up to order four
in the naive scaling can be constructed as follows:
\begin{align}
\mathcal{L}_{E}  &
=\mathop{\mathrm{Tr}}[(D_{\mu}\Phi_E)^{\dagger}(D_{\mu}\Phi_E)]+\alpha
\mathop{\mathrm{Tr}}[(D_{\mu}\Phi_E)^{\dagger}(D_{\mu}\Phi) + (D_{\mu}\Phi)^{\dagger}(D_{\mu}\Phi_E)]-(m_{0}%
^{\ast})^{2}\left(  \frac{G}{G_{0}}\right)  ^{2}\mathop{\mathrm{Tr}}(\Phi
_{E}^{\dagger}\Phi_{E})\nonumber\\
&  -\lambda_{0}\left(  \frac{G}{G_{0}}\right)  ^{2}\mathop{\mathrm{Tr}}(\Phi
_{E}^{\dagger}\Phi+\Phi^{\dagger}\Phi_{E})-\lambda_{1}^{\ast}%
\mathop{\mathrm{Tr}}(\Phi_{E}^{\dagger}\Phi_{E})\mathop{\mathrm{Tr}}(\Phi
^{\dagger}\Phi)-\lambda_{2}^{\ast}\mathop{\mathrm{Tr}}(\Phi_{E}^{\dagger}%
\Phi_{E}\Phi^{\dagger}\Phi+\Phi_{E}\Phi_{E}^{\dagger}\Phi\Phi^{\dagger
})\nonumber\\
&  -\kappa_{1}\mathop{\mathrm{Tr}}(\Phi_{E}^{\dagger}\Phi+\Phi^{\dagger}%
\Phi_{E})\mathop{\mathrm{Tr}}(\Phi^{\dagger}\Phi)-\kappa_{2}%
[\mathop{\mathrm{Tr}}(\Phi_{E}^{\dagger}\Phi+\Phi^{\dagger}\Phi_{E}%
)]^{2}-\kappa_{3}\mathop{\mathrm{Tr}}(\Phi_{E}^{\dagger}\Phi+\Phi^{\dagger
}\Phi_{E})\mathop{\mathrm{Tr}}(\Phi_{E}^{\dagger}\Phi_{E})-\kappa
_{4}[\mathop{\mathrm{Tr}}(\Phi_{E}^{\dagger}\Phi_{E})]^{2}\nonumber\\
&  -\xi_{1}\mathop{\mathrm{Tr}}(\Phi_{E}^{\dagger}\Phi\Phi^{\dagger}\Phi
+\Phi_{E}\Phi^{\dagger}\Phi\Phi^{\dagger})-\xi_{2}\mathop{\mathrm{Tr}}(\Phi
_{E}^{\dagger}\Phi\Phi_{E}^{\dagger}\Phi+\Phi^{\dagger}\Phi_{E}\Phi^{\dagger
}\Phi_{E})-\xi_{3}\mathop{\mathrm{Tr}}(\Phi^{\dagger}\Phi_{E}\Phi_{E}%
^{\dagger}\Phi_{E}+\Phi\Phi_{E}^{\dagger}\Phi_{E}\Phi_{E}^{\dagger})-\xi
_{4}\mathop{\mathrm{Tr}}(\Phi_{E}^{\dagger}\Phi_{E})^{2}{\nonumber}\nonumber\\
&  +\mathop{\mathrm{Tr}}(\Phi_{E}^{\dagger}\Phi_{E}E_{1}+\Phi_{E}\Phi
_{E}^{\dagger}E_{1})+c_{1}^{\ast}[(\det\Phi-\det\Phi_{E}^{\dagger})^{2}%
+(\det\Phi^{\dagger}-\det\Phi_{E})^{2}]+c_{1E}^{\ast}(\det\Phi_{E}-\det
\Phi_{E}^{\dagger})^{2}{\nonumber}\nonumber\\
&  +\frac{h_{1}^{\ast}}{2}\mathop{\mathrm{Tr}}(\Phi_{E}^{\dagger}\Phi
+\Phi^{\dagger}\Phi_{E})\mathop{\mathrm{Tr}}(L_{\mu}^{2}+R_{\mu}^{2}%
)+\frac{h_{1E}^{\ast}}{2}\mathop{\mathrm{Tr}}(\Phi_{E}^{\dagger}\Phi
_{E})\mathop{\mathrm{Tr}}(L_{\mu}^{2}+R_{\mu}^{2}){\nonumber}\nonumber\\
&  +h_{2}^{\ast}\mathop{\mathrm{Tr}}(\Phi_{E}^{\dagger}L_{\mu}L^{\mu}\Phi
+\Phi^{\dagger}L_{\mu}L^{\mu}\Phi_{E}+R_{\mu}\Phi_{E}^{\dagger}\Phi R^{\mu
}+R_{\mu}\Phi^{\dagger}\Phi_{E}R^{\mu})+h_{2E}^{\ast}%
\mathop{\mathrm{Tr}} [\vert L_{\mu}\Phi_E \vert ^{2}+\vert \Phi_E R_{\mu} \vert ^{2}]{\nonumber}\nonumber\\
&  +2h_{3}^{\ast}\mathop{\mathrm{Tr}}(L_{\mu}\Phi_{E}R^{\mu}\Phi^{\dagger
}+L_{\mu}\Phi R^{\mu}\Phi_{E}^{\dagger})+2h_{3E}^{\ast}%
\mathop{\mathrm{Tr}}(L_{\mu}\Phi_{E}R^{\mu}\Phi_{E}^{\dagger})\text{ .}
\label{LagrangianE}%
\end{align}
\\
The particle content of the Lagrangian is the same as the one in
Eqs.\ (\ref{eq:matrix_field_L}) and (\ref{eq:matrix_field_R}) for spin-1
states and it is analogous to Eq.\ (\ref{eq:matrix_field_Phi}) for
(pseudo)scalars:
\begin{equation}
\Phi_{E}=\frac{1}{\sqrt{2}}\left(
\begin{array}
[c]{ccc}%
\frac{(\sigma_{N}^{E}+a_{0}^{0E})+i(\eta_{N}^{E}+\pi^{0E})}{\sqrt{2}} &
a_{0}^{+E}+i\pi^{+E} & K_{0}^{\star+E}+iK^{+E}\\
a_{0}^{-E}+i\pi^{-E} & \frac{(\sigma_{N}^{E}-a_{0}^{0E})+i(\eta_{N}^{E}%
-\pi^{0E})}{\sqrt{2}} & K_{0}^{\star0E}+iK^{0E}\\
K_{0}^{\star-E}+iK^{-E} & {\bar{K}_{0}^{\star0E}}+i{\bar{K}^{0E}} & \sigma
_{S}^{E}+i\eta_{S}^{E}%
\end{array}
\right)  \text{ .} \label{PhiE}%
\end{equation}
The covariant derivative $D^{\mu}\Phi^{E}$ is defined analogously to
Eq.\ (\ref{Dmu}):
\begin{equation}
D^{\mu}\Phi^{E}\equiv\partial^{\mu}\Phi^{E}-ig_{1}^{E}(L^{\mu}\Phi^{E}%
-\Phi^{E}R^{\mu})
\label{DmuE}%
\end{equation}
and we also set $E_{1}=$ diag$\{0,0,\epsilon_{S}^{E}\}$.
\\
\\
Spontaneous symmetry breaking in the Lagrangian for the excited
(pseudo)scalars will be implemented only by means of condensation of
ground-state quarkonia $\sigma_{N}$ and $\sigma_{S}$, i.e., as a first
approximation, we assume that their excited counterparts $\sigma_{N}^{E}$ and
$\sigma_{S}^{E}$ do not condense\footnote{There is a subtle point pertaining
to the condensation of excited states in $\sigma$-type models: as discussed in
Ref.\ \cite{Francesco2}, it can be in agreement with QCD constraints but may
also, depending on parameter choice, spontaneously break parity in vacuum.
Study of a model with condensation of the excited states
would go beyond the current work. (It would additionally imply that the 
excited pseudoscalars also represent Goldstone bosons of QCD which is disputed in, e.g., 
Ref.\ \cite{Holl:2004fr}.)}. As a consequence,
there is no need to shift spin-1 fields or renormalise the excited
pseudoscalars as described in Eqs.\ (\ref{Vmu}) - (\ref{ZetaS}).
\\
\\
We now turn to the assignment of the excited states. Considering isospin
multiplets as single degrees of freedom, there are 8 states in
Eq.\ (\ref{PhiE}): $\sigma_{N}^{E}$, $\sigma_{S}^{E}$, $\vec{a}_{0}^{E}$ and
$K_{0}^{\star E}$ (scalar) and $\eta_{N}^{E}$, $\eta_{S}^{E}$, $\vec{\pi}^{E}$
and $K^{E}$ (pseudoscalar); the
experimental information on states with these quantum numbers is at times
limited or their identification is unclear:

\begin{itemize}
\item Seven states are listed by the PDG in the scalar isosinglet ($IJ^{PC}%
=00^{++}$) channel in the energy region up to $\simeq$ 2 GeV: $f_{0}(500)$/$\sigma$,
$f_{0}(980)$, $f_{0}(1370)$, $f_{0}(1500)$, $f_{0}(1710)$, $f_{0}(2020)$ 
and $f_0(2100)$.
The last two are termed as unestablished \cite{PDG}; the others have been
subject of various studies in the last decades
\cite{Article2010,Article2013,Basdevant:1972uu,Estabrooks:1978de,Close1988,vanBeveren:1986ea,Zou:1994ea,
Kaminski:1993zb,Achasov:1994iu,Tornqvist:1995,Muenz:1996,Dobado:1996ps,Elias:1998bq,
Black:1998wt,Minkowski:1998mf,Oller:1998zr,Kaminski:1998ns,Ishida:1999qk,Surovtsev:2000ev,
Black:2000qq,Teshima2002,scalars-above1GeVqq-below1GeVq2q2,Anisovich-KMatrix,
Pelaez2003,Bugg:2003kj,Scadron:2003yg,Napsuciale2004,Pelaez:2004xp,Lattice3,
Lattice5,Fariborz:2007ai,Albaladejo:2008qa,Fariborz2009,Mennessier:2010xg,
Branz:2010gd,GarciaMartin:2011jx,Mukherjee:2012xn,Fariborz:2015era,
Eichmann:2015cra,Pelaez:2015qba,Fariborz:2015dou,Kirchbach:2016scz}. As mentioned in the Introduction, the
general conclusion is that the states up to and including $f_{0}(1710)$ are
compatible with having ground-state $\bar{q}q$ or $\bar{q}\bar{q}qq$
structure; the presence of the scalar glueball is also expected
\cite{Stani1,Stani2,Albaladejo:2008qa,Amsler1995,LatticeGlueball1,Lee1999,Close2001,
Amsler2004,LatticeGlueball2,Close2005,Giacosa2005,LatticeGlueball3,Forkel2007,
Mathieu2008,LatticeGlueball4,LatticeGlueball5,Brunner:2015oqa,Brunner:2015yha,Capossoli1,
Brunner:2015oga,Capossoli2,Capossoli3,Capossoli4}. However, none of
these states is considered as the first radial excitation of the scalar
isosinglet $\bar{q}q$ state.\newline A decade ago, a new resonance named
$f_{0}(1790)$ was observed by the BES II Collaboration in the $\pi\pi$ final states
produced in $J/\Psi$ radiative decays \cite{BESII}; there had been
evidence for this state in the earlier data of MARK III \cite{Bugg:1995jq} and BES \cite{Bai:1999mm}. 
Recently, LHCb has confirmed this finding in a study of $B_{s}\rightarrow J/\Psi\pi\pi$ decays
\cite{LHCb}. Since, as indicated, the spectrum of
ground-state scalar quarkonia appears to be contained in the already established
resonances, we will work here with the hypothesis that $f_{0}(1790)$ is the
first excitation of the $\bar{n}n$ ground state ($\equiv\sigma_{N}^{E}$).
The assignment is further motivated by the predominant coupling
of $f_0(1790)$ to pions \cite{BESII}.
\newline The data of Ref.\ \cite{BESII} will be used as follows: $m_{f_{0}%
(1790)}=(1790\pm35)$ MeV and $\Gamma_{f_{0}(1790)\rightarrow\pi\pi}%
=(270\pm45)$ MeV, with both errors made symmetric and given as arithmetic means of those published by
BES\ II. Additionally, Ref.\ \cite{BESII} also reports the branching ratios
$J/\Psi\rightarrow\phi f_{0}(1790)\rightarrow\phi\pi\pi=(6.2\pm1.4)\cdot
10^{-4}$ and $J/\Psi\rightarrow\phi f_{0}(1790)\rightarrow\phi KK=(1.6\pm
0.8)\cdot10^{-4}$. Using $\Gamma_{f_{0}(1790)\rightarrow\pi\pi}=(270\pm45)$
MeV and the quotient of the mentioned branching ratios we estimate
$\Gamma_{f_{0}(1790)\rightarrow KK}=(70\pm40)$ MeV. These data will become 
necessary in Sections
\ref{Masses2} and \ref{Decays2}. We note, however, already at this point that the large uncertainties
in $f_0(1790)$ decays -- a direct consequence of uncertainties 
in the $J/\Psi$ branching ratios amounting to $\sim$23\% and 50\% -- will lead to ambiguities in prediction of some
decays (see Sec.\ \ref{Decays3}). These are nonetheless the most comprehensive data available at the moment,
and more data would obviously be of great importance.
\newline The
assignment of our excited isoscalar $\bar{s}s$ state $\sigma_{S}^{E}$ will be
discussed as a consequence of the model [particularly in the context of
$f_{0}(2020)$ and $f_0(2100)$].

\item Two resonances are denoted as established by the PDG in the
$IJ^{PC}=10^{++}$ channel: $a_{0}(980)$ and $a_{0}(1450)$ \cite{PDG}. Various
interpretations of these two states in terms of ground-state $\bar{q}q$ or
$\bar{q}\bar{q}qq$ structures or meson-meson molecules have been proposed
\cite{Tornqvist:1995,Elias:1998bq,Black:1998wt,Oller:1998zr,Ishida:1999qk,
Black:2000qq,Teshima2002,scalars-above1GeVqq-below1GeVq2q2,Scadron:2003yg,Napsuciale2004,
Pelaez:2004xp,Fariborz:2007ai,Fariborz2009,Eichmann:2015cra,Kirchbach:2016scz,
Lattice3,Lattice5,Lattice1,Lattice2,Lattice6}.\newline Recently, the BABAR
Collaboration \cite{BABAR} has claimed the observation of a new resonance
denoted as $a_{0}(1950)$ in the process $\gamma\gamma\rightarrow\eta
_{c}(1S)\rightarrow\bar{K}K\pi$ with significance up to 4.2 $\sigma$.
There was earlier evidence for this state in the Crystal Barrel data \cite{Anisovich:1999jv,Bugg:2004xu}; see also Refs.\ \cite{Afonin:2007jd,Afonin:2007aa}.
We will
discuss the possible interpretation of this resonance in terms of the first
$IJ^{PC}=10^{++}$excitation as a result of our calculations.

\item Two resonances are candidates for the ground-state $\bar{q}q$ resonance in
the scalar-kaon channel (with alternative interpretations -- just as in the
case of the $a_{0}$ resonances -- in terms of $\bar{q}\bar{q}qq$ structures or
meson-meson molecules): $K_{0}^{\star}(800)$/$\kappa$ and $K_{0}^{\star
}(1430)$; controversy still surrounds the first of these states
\cite{Estabrooks:1978de,Tornqvist:1995,Oller:1998zr,Ishida:1999qk,Black:2000qq,
Teshima2002,scalars-above1GeVqq-below1GeVq2q2,Pelaez2003,Bugg:2003kj,Napsuciale2004,Lattice3,
Eichmann:2015cra,Lattice2,Lattice4,Lattice6}. \newline A possibility is that
$K_{0}^{\star}(1950)$, the highest-lying resonance in this channel, represents
the first excitation, although the state is (currently) unestablished
\cite{PDG}. This will be discussed as a result of our calculations later on.

\item The pseudoscalar isosinglet ($IJ^{PC}=00^{-+}$) channel has six known
resonances in the energy region below 2 GeV according to the PDG \cite{PDG}:
$\eta$, $\eta^{\prime}(958)$, $\eta(1295)$, $\eta(1405)$, $\eta(1475)$ and
$\eta(1760)$. \newline Not all of them are without controversy: for example,
the observation of $\eta(1405)$ and $\eta(1475)$ as two different states was
reported by E769 \cite{BNL1989}, E852 \cite{BNL2001}, MARK III
\cite{MarkIII19990}, DM2 \cite{DM21992} and OBELIX
\cite{OBELIX1999,OBELIX2002} while they were claimed to represent a single
state named $\eta(1440)$ by Crystal Ball \cite{CB1983} and BES
\cite{BES1998,BES2000} Collaborations. It is important to note that a clear
identification of pseudoscalar resonance(s) in the energy region between 1.4
GeV and 1.5 GeV depends strongly on a proper consideration, among other, of
the $K^{\star}K$ threshold opening ($m_{K^{\star}}+m_{K}=1385$ MeV) and of the
existence of the $IJ^{PC}=01^{++}$ state $f_{1}(1420)$ whose partial wave is
known to influence the pseudoscalar one in experimental analyses (see, e.g.,
Ref.\ \cite{BNL2001}). A comprehensive study of BES II data in
Ref.\ \cite{BuggBESII}, that included an energy-dependent Breit-Wigner
amplitude as well as a dispersive correction to the Breit-Wigner denominator
(made necessary by the proximity to the $K^{\star}K$ threshold), has observed
only a marginal increase in fit quality when two pseudoscalars are considered.
In line with this, our study will assume the existence of $\eta(1440)$ to
which our $\eta_{S}^{E}$ state will be assigned. We will use $m_{\eta
(1440)}=(1432\pm10)$ MeV and $\Gamma_{\eta(1440) \rightarrow
K^{\star} K} = (26 \pm 3)$ MeV 
\cite{BES1998,BES2000} in Sections \ref{Masses2} and \ref{Decays4}; 
the error in the decay width is our estimate.
We emphasise, however, that
our results are stable up to a $\lesssim$ 3\% change when $\eta(1475)$ is
considered instead of $\eta(1440)$ \footnote{The $\eta(1405)$ resonance would then be a
candidate for the pseudoscalar glueball \cite{Faddeev:2003aw}.}.
\newline Our
state $\eta_{N}^{E}$ will be assigned to $\eta(1295)$ in order to test the
hypothesis whether an excited pseudoscalar isosinglet at $\simeq1.3$ GeV can
be accommodated in eLSM (and notwithstanding the experimental concerns raised in
Ref.\ \cite{Klempt}). We will use the PDG value $m_{\eta(1295)}=(1294\pm4)$ MeV
for determination of mass parameters in Sec.\ \ref{Masses2}. The PDG also reports
$\Gamma_{\eta(1295)}^{\text{total}} = (55 \pm 5)$ MeV; the relative contributions of 
$\eta(1295)$ decay channels are uncertain. Nonetheless, we will use 
$\Gamma_{\eta(1295)}^{\text{total}}$ in Sec.\ \ref{Decays4}.

\item Two states have the quantum number of a pion excitation: $\pi(1300)$ and
$\pi(1800)$, with the latter being a candidate for a non-$\bar{q}q$ state
\cite{PDG}. The remaining $\pi(1300)$ resonance may in principle be an excited
$\bar{q}q$ isotriplet; however, due to the experimental uncertainties reported
by the PDG [$m_{\pi(1300)}=(1300\pm100)$ MeV but merely an interval for $\Gamma
_{\pi(1300)}=(200-600)$ MeV] this will only be discussed as a possible result
of our model.

\item Two states are candidates for the excited kaon: $K(1460)$ and $K(1830)$.
Since other excited states of our model have been assigned to resonances with
energies $\simeq$ 1.4 GeV, we will study the possibility that our
$IJ^{P}=\frac{1}{2}0^{-}$ state corresponds to $K(1460)$. This will, however,
only be discussed as a possible result of the model since the experimental
data on this state is very limited: $m_{K(1460)}\sim1460$ MeV; $\Gamma
_{K(1460)}\sim260$ MeV \cite{PDG}.
\end{itemize}
$\,$\\
As indicated in the above points, with regard to the use of the above data
for parameter determination we exclude as input all states for which there are only
scarce/unestablished data and, additionally, those for which the PDG cites only intervals
for mass/decay width (since the latter lead to weak parameter constraints). Then we are left
with only three resonances whose experimental data shall be used: $f_0(1790)$, $\eta(1295)$ and $\eta(1440)$.
For clarity, we collect the assignment of the model states (where possible), and also the data that we will use,
in Table \ref{Assignments}. The data are used in Sec.\ \ref{Results}.

\begin{table}[th]
\centering%
\begin{tabular}
[c]{|c|c|c|c|c|}\hline
Model state & $IJ^{P}$ & Assignment & We use \\\hline
$\sigma_{N}^{E}$ & $00^+$ & $f_{0}(1790)$ & $m_{f_0(1790)} = (1790 \pm 35)$ MeV \cite{BESII} \\
& & & $\Gamma_{f_{0}(1790)\rightarrow\pi\pi}=(270\pm45)$ MeV \cite{BESII} \\
& & & $\Gamma_{f_{0}(1790)\rightarrow KK}=(70\pm40)$ MeV \\\hline
$\eta_{N}^{E}$ & $00^-$ & $\eta(1295)$ & $m_{\eta(1295)}=(1294\pm4)$ MeV \cite{PDG} \\
& & & $\Gamma_{\eta(1295)}^{\text{total}} = (55 \pm 5)$ MeV \cite{PDG} \\\hline
$\eta_{S}^{E}$ & $00^-$ & $\eta(1440)$ & $m_{\eta(1440)}=(1432\pm10)$ MeV \cite{BES1998,BES2000} \\
& & & $\Gamma_{\eta(1440) \rightarrow K^{\star} K} = (26 \pm 3)$ MeV \\\hline
$\sigma_{S}^{E}$ & $00^+$ & Possible overlap with $f_{0}(2020)$/$f_0(2100)$ & \\
&    & to be discussed as a model consequence & - \\\hline
$a_{0}^{E}$ & $10^+$ & Possible overlap with $a_{0}(1950)$ & \\
&   & to be discussed as a model consequence & - \\\hline
$\pi^{E}$ & $10^-$ & Possible overlap with $\pi(1300)$ & \\
&    & to be discussed as a model consequence & - \\\hline
$K_{0}^{\star E}$ & $\frac{1}{2} 0^+$ & Possible overlap with $K_{0}^{\star}(1950)$ & \\
&    & to be discussed as a model consequence & - \\\hline
$K^{E}$ & $\frac{1}{2} 0^-$ & Possible overlap with $K(1460)$ & \\
&   & to be discussed as a model consequence & - \\\hline
\end{tabular}
\caption{Assignment of the states in Eq.\ (\ref{PhiE}) to physical states. Note:
every assignment implies the hypothesis that the physical state has the
$\bar{q}q$ structure.}%
\label{Assignments}%
\end{table}

\subsubsection{Parameters}

\label{Parameters1}

The following parameters are present in Eq.\ (\ref{LagrangianE}):%
\begin{equation}
g_{1}^{E}\text{, }\alpha\text{, }m_{0}^{\ast}\text{, }\lambda_{0}\text{,
}\lambda_{1,2}^{\ast}\text{, }\kappa_{1,2,3,4}\text{, }\xi_{1,2,3,4}\text{,
}\epsilon_{S}^{E}\text{, }c_{1}^{\ast}\text{, }c_{1}^{\ast E}\text{,
}h_{1,2,3}^{\ast}\text{, }h_{1,2,3}^{\ast E}\;.
\end{equation}
$\,$
\\
The number of parameters relevant for masses and decays of the excited states
is significantly smaller as apparent once the following selection criteria are applied:

\begin{itemize}
\item All large-$N_{c}$ suppressed parameters are set to zero since their
influence on the general phenomenology is expected to be small and the current
experimental uncertainties do not permit their determination. Hence the
parameters $\lambda_{1}^{\ast}$, $h_{1}^{\ast}$ and $\kappa_{1,2,3,4}$ are discarded.

\item The parameter $c_{1}^{\ast}$ is set to zero since it contains a term
$\sim(\det\Phi)^{2}$ that would influence ground-state mass terms after
condensation of $\sigma_{N}$ and $\sigma_{S}$. Such introduction of an
additional parameter is not necessary since, as demonstrated in
Ref.\ \cite{Article2013}, the ground states are very well described by
Lagrangian (\ref{Lagrangian}).

\item As a first approximation, we will discard all parameters that lead to
particle mixing and study whether the assignments described in Table
\ref{Assignments} are compatible with experiment. Hence we discard the
parameters $\alpha$, $\lambda_{0}$ and $\xi_{1}$; note that mixing is also
induced by $\kappa_{1,2}$ and $c_{1}^{\ast}$ but these have already been
discarded for reasons stated above\footnote{However, there would be no mixing
of pseudoscalar isosinglets $\eta_{N}^{E}$ and $\eta_{S}^{E}$ in the model
even if all discarded parameters were considered. The reason is that there is
no condensation of excited scalar states in Lagrangian (\ref{LagrangianE}).}.

\item Parameters that lead to decays with two or more excited final states are
not of relevance for us: all states in the model have masses between $\sim1$
GeV and $\sim2$ GeV and hence such decays are kinematically forbidden.
(Parameters $\lambda_{2}^{\ast}$ and $\xi_{2}$\ that contribute to mass terms
are obviously relevant and excepted from this criterion.) Hence we can discard
$\xi_{3,4}$, $c_{1}^{\ast E}$ and $h_{1,2,3}^{\ast E}$.
\end{itemize}
Note that the above criteria are not mutually exclusive: some parameters may
be set to zero on several grounds, such as for example $\kappa_{1}$.
\\
\\
Consequently we are left with the following undetermined parameters:%
\begin{equation}
g_{1}^{E}\text{, }m_{0}^{\ast}\text{, }\lambda_{2}^{\ast}\text{, }\xi
_{2}\text{, }\epsilon_{S}^{E}\text{, }h_{2,3}^{\ast}\;.
\end{equation}
$\,$
\\
The number of parameters that we will actually use is even smaller, as we
discuss in Sections \ref{Masses1} and \ref{Decays1}.

\subsubsection{Mass Terms}

\label{Masses1}

The following mass terms are obtained for the excited states present in the model:%

\begin{align}
m_{\sigma_{N}^{E}}^{2}  &  =(m_{0}^{\ast})^{2}+\frac{\lambda_{2}^{\ast}%
+\xi_{2}}{2}\phi_{N}^{2}\label{sigmaNE}\\
m_{a_{0}^{E}}^{2}  &  =(m_{0}^{\ast})^{2}+\frac{\lambda_{2}^{\ast}+\xi_{2}}%
{2}\phi_{N}^{2}\\
m_{\pi^{E}}^{2}  &  =m_{\eta_{N}^{E}}^{2}=(m_{0}^{\ast})^{2}+\frac{\lambda
_{2}^{\ast}-\xi_{2}}{2}\phi_{N}^{2}\label{etaNE}\\
m_{\eta_{S}^{E}}^{2}  &  =(m_{0}^{\ast})^{2}-2\epsilon_{S}^{E}+\left(
\lambda_{2}^{\ast}-\xi_{2}\right)  \phi_{S}^{2}\label{etaSE}\\
m_{\sigma_{S}^{E}}^{2}  &  =(m_{0}^{\ast})^{2}-2\epsilon_{S}^{E}+\left(
\lambda_{2}^{\ast}+\xi_{2}\right)  \phi_{S}^{2}\\
m_{K^{E}}^{2}  &  =(m_{0}^{\ast})^{2}-\epsilon_{S}^{E}+\frac{\lambda_{2}%
^{\ast}}{4}\phi_{N}^{2}-\frac{\xi_{2}}{\sqrt{2}}\phi_{N}\phi_{S}+\frac
{\lambda_{2}^{\ast}}{2}\phi_{S}^{2}\label{KE}\\
m_{K_{0}^{\star E}}^{2}  &  =(m_{0}^{\ast})^{2}-\epsilon_{S}^{E}+\frac
{\lambda_{2}^{\ast}}{4}\phi_{N}^{2}+\frac{\xi_{2}}{\sqrt{2}}\phi_{N}\phi
_{S}+\frac{\lambda_{2}^{\ast}}{2}\phi_{S}^{2}\;. \label{KSE}%
\end{align}
$\;$
\\
The mass terms (\ref{sigmaNE}) - (\ref{KSE}) contain the same linear
combination of $m_{0}^{\ast}$ and $\lambda_{2}^{\ast}$:%
\begin{equation}
C_{1}^{\ast}=(m_{0}^{\ast})^{2}+\frac{\lambda_{2}^{\ast}}{2}\phi_{N}^{2}
\label{C1star}%
\end{equation}
$\;$
\\
and the mass terms (\ref{etaSE}) - (\ref{KSE}) contain the same linear
combination of $\lambda_{2}^{\ast}$ and $\epsilon_{S}^{E}$:
\begin{equation}
C_{2}^{\ast}=\lambda_{2}^{\ast}Z_{K}f_{K}(Z_{K}f_{K}-\phi_{N})-\epsilon
_{S}^{E}\;. \label{C2star}%
\end{equation}
This is obvious after substituting the strange condensate $\phi_{S}$ by the
non-strange condensate $\phi_{N}$ via Eq.\ (\ref{PhiS}). The modified mass
terms then read%
\begin{align}
m_{\sigma_{N}^{E}}^{2}  &  =C_{1}^{\ast}+\frac{\xi_{2}}{2}\phi_{N}%
^{2}\label{sigmaNE1}\\
m_{a_{0}^{E}}^{2}  &  =C_{1}^{\ast}+\frac{\xi_{2}}{2}\phi_{N}^{2} \label{a0E1}\\
m_{\pi^{E}}^{2}  &  =m_{\eta_{N}^{E}}^{2}=C_{1}^{\ast}-\frac{\xi_{2}}{2}%
\phi_{N}^{2} \label{etaNE1}\\
m_{\eta_{S}^{E}}^{2}  &  =C_{1}^{\ast}+2C_{2}^{\ast}-\frac{\xi_{2}}{2}%
(\phi_{N}-2Z_{K}f_{K})^{2} \label{etaSE1}\\
m_{\sigma_{S}^{E}}^{2}  &  =C_{1}^{\ast}+2C_{2}^{\ast}+\frac{\xi_{2}}{2}%
(\phi_{N}-2Z_{K}f_{K})^{2} \label{sigmaSE1}\\
m_{K^{E}}^{2}  &  =C_{1}^{\ast}+C_{2}^{\ast}+\frac{\xi_{2}}{2}\phi_{N}%
(\phi_{N}-2Z_{K}f_{K}) \label{KE1}\\
m_{K_{0}^{\star\,E}}^{2}  &  =C_{1}^{\ast}+C_{2}^{\ast}-\frac{\xi_{2}}{2}%
\phi_{N}(\phi_{N}-2Z_{K}f_{K})\;. \label{KSE1}%
\end{align}
Mass terms for all eight excited states can hence be described in terms of
only three parameters from Eq.\ (\ref{LagrangianE}): $C_{1}^{\ast}$,
$C_{2}^{\ast}$ and $\xi_{2}$.

\subsubsection{Decay Widths}

\label{Decays1}

Our objective is to perform a tree-level calculation of all kinematically
allowed two- and three-body decays for all excited states present in the
model. The corresponding interaction Lagrangians are presented in Appendix
\ref{Appendix}. As we will see, there are more than 35 decays that can be determined in this
way but all of them can be calculated using only a few formulas.
\\
\\
The generic formula for the decay width of particle $A$ into particles $B$ and
$C$ reads%

\begin{equation}
\Gamma_{A\rightarrow BC}=\mathcal{I}\frac{|\mathbf{k}|}{8\pi m_{A}^{2}%
}\left\vert \mathcal{M}_{A\rightarrow BC}\right\vert ^{2}\;, \label{Gamma2}%
\end{equation}
$\;$
\\
where $\mathbf{k}$ is the three-momentum of one of the final states in the
rest frame of $A$ and $\mathcal{M}$ is the decay amplitude (i.e., transition
matrix element). $\mathcal{I}$ is a symmetry factor emerging from the isospin
symmetry -- it is determined by the number of sub-channels for a given set of
final states (e.g., $\mathcal{I}$ $=2$ if $B$ and $C$ both correspond to
kaons). Usual symmetry factors are included if the final states are identical.
As we will see in Sec.\ \ref{Decays2}, decay widths obtained in the model are generally 
much smaller than resonance masses; for this reason, we do not expect large 
unitarisation effects \cite{Pagliara}.
\\
\\
Depending on the final states, the interaction Lagrangians presented in
Appendix \ref{Appendix} can have one of the following general structures:

\begin{itemize}
\item For a decay of the form $S\rightarrow P_{1}P_{2}$, where $S$ is a scalar
and $P_{1}$ and $P_{2}$ are pseudoscalar particles, the generic structure of
the interaction Lagrangian is%
\begin{equation}
\mathcal{L}_{SP_{1}P_{2}}=D_{SP_{1}P_{2}}\,SP_{1}P_{2}+E_{SP_{1}P_{2}%
}\,S\partial_{\mu}P_{1}\partial^{\mu}P_{2}+F_{SP_{1}P_{2}}\,\partial_{\mu
}S\partial^{\mu}P_{1}P_{2}\;,
\end{equation}

where $D_{SP_{1}P_{2}}$, $E_{SP_{1}P_{2}}$ and $F_{SP_{1}P_{2}}$ are
combinations of (some of the) parameters entering Lagrangian
(\ref{LagrangianE}). According to Eq.\ (\ref{Gamma2}), the decay width reads
in this case
\begin{equation}
\Gamma_{S\rightarrow P_{1}P_{2}}=\mathcal{I}\frac{|\mathbf{k}|}{8\pi m_{S}%
^{2}}\left\vert D_{SP_{1}P_{2}}-E_{SP_{1}P_{2}}\,K_{1}\cdot K_{2}%
+F_{SP_{1}P_{2}}\,K\cdot K_{1}\right\vert ^{2}\;, \label{SPP}%
\end{equation}

where $K$, $K_{1}$ and $K_{2}$ are respectively 4-momenta of $S$, $P_{1}$ and
$P_{2}$.

\item For a decay of the form $S\rightarrow VP$, where $V$ is a vector and $P$
is a pseudoscalar particle, the generic structure of the interaction
Lagrangian is%
\begin{equation}
\mathcal{L}_{SVP}=D_{SVP}\,SV_{\mu}\partial^{\mu}P\;,
\end{equation}

where $D_{SVP}$ is a combination of (some of the) parameters entering
Lagrangian (\ref{LagrangianE}). The decay width reads in this case
\begin{equation}
\Gamma_{S\rightarrow VP}=\mathcal{I}\frac{|\mathbf{k}|}{8\pi m_{S}^{2}}%
D_{SVP}^{2}\left[  \frac{(m_{S}^{2}-m_{V}^{2}-m_{P}^{2})^{2}}{4m_{V}^{2}%
}-m_{P}^{2}\right]  \;. \label{SVP}%
\end{equation}

\item For a decay of the form $S\rightarrow V_{1}V_{2}$, where $V_{1}$ and
$V_{2}$ are vector particles, the generic structure of the interaction
Lagrangian is%
\begin{equation}
\mathcal{L}_{SV_{1}V_{2}}=D_{SV_{1}V_{2}}\,SV_{1\mu}V_{2}^{\mu}\;,
\end{equation}

where $D_{SV_{1}V_{2}}$ is a combination of (some of the) parameters entering
Lagrangian (\ref{LagrangianE}). Then the decay width reads%
\begin{equation}
\Gamma_{S\rightarrow V_{1}V_{2}}=\mathcal{I}\frac{|\mathbf{k}|}{4\pi m_{S}%
^{2}}D_{SV_{1}V_{2}}^{2}\left[  \frac{(m_{S}^{2}-m_{V_{1}}^{2}-m_{V_{2}}%
^{2})^{2}}{8m_{V_{1}}^{2}m_{V_{2}}^{2}}+1\right]  \;. \label{SVV}%
\end{equation}

\end{itemize}
As evident from Appendix \ref{Appendix}, the most general interaction
Lagrangian for 3-body decays of the form $S\rightarrow S_{1}S_{2}S_{3}$ is

\begin{align}
\mathcal{L}_{SS_{1}S_{2}S_{3}}  &  =D_{SS_{1}S_{2}S_{3}}\,SS_{1}S_{2}%
S_{3}+E_{SS_{1}S_{2}S_{3}}\,S(\partial_{\mu}S_{1}\partial^{\mu}S_{2}%
)S_{3}\nonumber\\
&  +(\text{analogous terms with derivative couplings among final states
only})\;.
\end{align}
\\
The ensuing formula for the decay width reads%
\begin{equation}
\Gamma_{S\rightarrow S_{1}S_{2}S_{3}}=\mathcal{I}\frac{1}{32(2\pi)^{3}%
m_{S}^{3}}\int_{(m_{S_{1}}+m_{S_{2}})^{2}}^{(m_{S}-m_{S_{3}})^{2}}%
\,\text{d}m_{12}^{2}\int_{(m_{23})_{\min.}}^{(m_{23})_{\max.}}\,\text{d}%
m_{23}^{2}\,\left\vert \mathcal{M}_{S\rightarrow S_{1}S_{2}S_{3}}\right\vert
^{2}\; \label{Gamma3}%
\end{equation}
\\
where $m_{12}^{2}=(K_{S_{1}}+K_{S_{2}})^{2}$, $m_{23}^{2}=(K_{S_{2}}+K_{S_{3}%
})^{2}$ and%

\begin{align}
(m_{23})_{\min.}  &  =(E_{2}^{\ast}+E_{3}^{\ast})^{2}-\left[  \sqrt
{(E_{2}^{\ast})^{2}-m_{S_{2}}^{2}}+\sqrt{(E_{3}^{\ast})^{2}-m_{S_{3}}^{2}%
}\right]  ^{2}\\
(m_{23})_{\max.}  &  =(E_{2}^{\ast}+E_{3}^{\ast})^{2}-\left[  \sqrt
{(E_{2}^{\ast})^{2}-m_{S_{2}}^{2}}-\sqrt{(E_{3}^{\ast})^{2}-m_{S_{3}}^{2}%
}\right]  ^{2}%
\end{align}
\\
with%
\begin{equation}
E_{2}^{\ast}=\frac{m_{12}^{2}-m_{S_{1}}^{2}+m_{S_{2}}^{2}}{2m_{12}}%
\;,\;E_{3}^{\ast}=\frac{m_{S}^{2}-m_{12}^{2}-m_{S_{3}}^{2}}{2m_{12}}\;.
\end{equation}
$\;$
\\
As evident from Appendix \ref{Appendix}, our decay widths depend on the
following parameters: $g_{1}^{E}$, $\lambda_{2}^{\ast}$, $\xi_{2}$ and
$h_{2,3}^{\ast}$. The first three appear only in decays with an excited final
state; since such decays are experimentally unknown, it is not possible to
determine these parameters (and $\xi_{2}$ can be determined from the mass
terms in any case, see Sec.\ \ref{Masses1}). The remaining two, $h_{2,3}%
^{\ast}$, can be calculated from decays with ground states in the outgoing
channels -- we will discuss this in Sec.\ \ref{Decays2}.

\section{Masses and Decays of the Excited States: Results and Consequences}
\label{Results}

\subsection{Parameter Determination: General Remarks}

\label{Parameters2}

Combining parameter discussion at the end of Sections \ref{Masses1} and
\ref{Decays1}, the final conclusion is that the following parameters need to
be determined:%
\begin{equation}
C_{1}^{\ast}\text{, }C_{2}^{\ast}\text{, }\xi_{2}\text{, }h_{2}^{\ast}\text{
and }h_{3}^{\ast} \label{Parametere}
\end{equation}
with $C_{1}^{\ast}$ and $C_{2}^{\ast}$ parameter combinations defined in
Eqs.\ (\ref{C1star}) and (\ref{C2star}).
\\
\\
As evident from mass terms (\ref{sigmaNE1}) - (\ref{KSE1}) and Appendix
\ref{Appendix}, $C_{1}^{\ast}$ and $C_{2}^{\ast}$ influence only masses;
$\xi_{2}$ appears in decays with one excited final state and in mass terms.
Since, as indicated at the end of Sec.\ \ref{Decays1}, decays with excited
final states are experimentall unknown, $\xi_{2}$ can only be determined from
the masses. Contrarily, $h_{2}^{\ast}$ and $h_{3}^{\ast}$ appear only in decay
widths (with no excited final states). Hence our parameters are divided in two
sets, one determined by masses ($C_{1}^{\ast}$, $C_{2}^{\ast}$ and $\xi_{2}$)
and another determined by decays (${h_{2}^{\star}}$ and ${h_{3}^{\star}}$).
\\
\\
Parameter determination will ensue by means of a $\chi^{2}$ fit. Scarcity of
experimental data compels us to have an equal number of parameters and
experimental data entering the fit; although in that case the equation systems
can also be solved exactly, an advantage of the $\chi^{2}$ fit is that error
calculation for parameters and observables is then straightforward.
\\
\\
The general structure of of the fit function $\chi^{2}$ fit is as follows:
\begin{equation}
\chi^{2}(p_{1},...,p_{m})=\sum_{i=1}^{n}\left(  \frac{O_{i}^{\text{th.}}%
(p_{1},...,p_{m})-O_{i}^{\text{exp.}}}{\Delta O_{i}^{\text{exp.}}}\right)
^{2}\label{chi2}%
\end{equation}
for a set of $n$\ (theoretical) observables $O_{i}^{\text{th.}}$ determined by
$m\leq n$ parameters $p_{j}$. In our case, $m=n=3$ for masses and $m=n=2$ for
decay widths. Central values and errors on the
experimental side are respectively denoted as $O_{i}^{\text{exp.}}$ and
$\Delta O_{i}^{\text{exp.}}$. Parameter errors $\Delta p_{i}$ are calculated
as the square roots of the diagonal elements of the inverse Hessian matrix
obtained from $\chi^{2}(p_{j})$. Theoretical errors $\Delta O_{i}$ for each
observable $O_{i}$ are calculated by diagonalising the Hesse matrix via a
special orthogonal matrix $M$%
\begin{equation}
MHM^{t}\equiv\text{diag}\{\text{eigenvalues of }H\}
\end{equation}
and rotating parameters $p_{i}$ such that%
\begin{equation}
\vec{q}=M(\vec{p}-\vec{p}_{\min.})
\end{equation}
where $\vec{p}$ contains all parameters and $\vec{p}_{\min.}$ realises the
minimum of $\chi^{2}(p_{1},...,p_{m})$. Then we can determine $\Delta O_{i}$
via
\begin{equation}
\Delta O_{i}=\sqrt{\sum_{j=1}^{n}\left(  \left.  \frac{\partial O_{i}%
(q_{1},...q_{m})}{\partial q_{j}}\right\vert _{\text{at fit value of }O_{i}%
}\Delta q_{j}\right)  ^{2}} \label{DeltaO}
\end{equation}
(see also Chapter 39 of the Particle Data Book \cite{PDG}).

\subsection{Masses of the Excited States} \label{Masses2}

Following the discussion of the experimental data on excited states in
Sec.\ \ref{LagrangianS} and particle assignment in Table \ref{Assignments}, we
use the following masses for the $\chi^{2}$\ fit of Eq.\ (\ref{chi2}):
$m_{\sigma_{N}^{E}}\equiv m_{f_{0}(1790)}=(1790\pm35)$ MeV, $m_{\eta_{N}^{E}%
}\equiv m_{\eta(1295)}=(1294\pm4)$ MeV and $m_{\eta_{S}^{E}}\equiv
m_{\eta(1440)}=(1432\pm10)$ MeV. Results for $C_{1}^{\ast}$, $C_{2}^{\ast}$
and $\xi_{2}$ are
\begin{align}
C_{1}^{\ast }& =(2.4\pm 0.6)\cdot 10^{6}\text{ [MeV}^{2}\text{], }%
C_{2}^{\ast }=(2.5\pm 0.2)\cdot 10^{5}\text{ [MeV}^{2}\text{], }\xi
_{2}=57\pm 5 \;. \label{Massparameters}
\end{align}
With these parameters, the general discussion from Sec.\ \ref{Parameters2}
allows us to immediately predict the masses of $\sigma_S^E$, $a_0^E$, $K_0^{\star E}$, $\pi^E$ and $K^E$.
They are presented in Table \ref{Masses}.

\begin{table}[th]
\centering%
\begin{tabular}
[c]{|c|c|c|c|}\hline
Model state & $IJ^{P}$ & Mass (MeV) & Note \\\hline
$\sigma_{N}^{E}$ & $00^+$ &  $1790 \pm 35$* & Assigned to $f_0(1790)$ \\\hline
$\eta_{N}^{E}$ & $00^-$ & $1294 \pm \, \,4$* & Assigned to $\eta(1295)$ \\\hline
$\eta_{S}^{E}$ & $00^-$ & $1432 \pm 10$* & Assigned to $\eta(1440)$ \\\hline
$\sigma_{S}^{E}$ & $00^+$ &  $1961 \pm 38$ & Possible overlap with $f_{0}(2020)$ or $f_0(2100)$  \\\hline
$ a_{0}^{E}$ & $10^+$ & $1790 \pm 35$ & Possible overlap with $a_{0}(1950)$  \\\hline
$K_{0}^{\star E}$ & $\frac{1}{2} 0^+$ & $1877 \pm 36$ & Possible overlap with $K_{0}^{\star}(1950)$ \\\hline
$\pi^{E}$ & $10^-$ &  $1294 \pm \, \, \,4$ & Possible overlap with $\pi(1300)$ \\\hline
$K^{E}$ & $\frac{1}{2} 0^-$ & $1366 \pm \, \, \, 6$ & Possible overlap with $K(1460)$ \\\hline
\end{tabular}
\caption{Masses of the excited states present in the model. Masses marked with an asterisk are used as input.
Note: there is mass degeneracy of $\sigma_N^E$ and $a_0^E$ because we have discarded large-$N_c$ suppressed 
parameters in our excited-state Lagrangian (\ref{LagrangianE}) -- see
Sec.\ \ref{Parameters1}. The degeneracy of $\eta_N^E$ and $\pi^E$ is a feature of the model.}%
\label{Masses}%
\end{table}

\subsection{Decays of the Excited States} \label{Decays2}

\subsubsection{Hypothesis: $f_0(1790)$ is an excited $\bar{q}q$ state} \label{Decays3}

We have concluded in Sec.\ \ref{Parameters2} that only two parameters are of relevance for all decays 
predictable in the model: $h_{2}^{\ast}$ and $h_{3}^{\ast}$. They can be determined from the data on
the $f_0(1790)$ resonance discussed in Sec.\ \ref{LagrangianS}: 
$\Gamma_{f_{0}(1790)\rightarrow\pi\pi}=(270\pm45)$ MeV and
$\Gamma_{f_{0}(1790)\rightarrow KK}=(70\pm40)$ MeV \cite{BESII}. 
Performing the $\chi^2$ fit described in Sec.\ \ref{Parameters2} we obtain the following parameter values:
\begin{align}
h_{2}^{\ast} = 67 \pm 63 \text{, } h_{3}^{\ast} = 79 \pm 63 \;. \label{h23*}
\end{align}
Large uncertainties for parameters are a consequence of propagation of the large errors for 
$\Gamma_{f_{0}(1790)\rightarrow\pi\pi}$ and particularly for $\Gamma_{f_{0}(1790)\rightarrow KK}$.
As described in Sec.\ \ref{LagrangianS}, $\Gamma_{f_{0}(1790)\rightarrow KK}$ was obtained as our estimate
relying upon $J/\Psi$
branching ratios reported by BES II \cite{BESII} that themselves had uncertainties between $\sim 23$\% and
50 \%. We emphasise, however, that such uncertainties do not necessarily have to translate into large errors
for the observables. The reason is that error claculation involves derivatives at central values of
parameters [see Eq.\ (\ref{DeltaO})]; small values of derivatives may then compensate the large parameter uncertainties. This is
indeed what we observe for most decays.
\\\\
There is a large number of decays that can be calculated using the
interaction Lagrangians in Appendix \ref{Appendix}, parameter values in Eq.\ (\ref{h23*}),
formulas for decay widths in Eqs.\ (\ref{SPP}), (\ref{SVP}), (\ref{SVV}) and
(\ref{Gamma3}) as well as Eq.\ (\ref{DeltaO}) for the errors of observables.
All results are presented in Table \ref{Decayst1}.

\begin{table}[th]
\centering%
\begin{tabular}{|c|c|c|c|c|@{}c@{}|}\hline
Model state
& 
$IJ^P$
& 
Mass (MeV)
&
Decay
&
Width (MeV)
&
Note  
\tabularnewline\hline

$\sigma_{N}^{E}$ & $00^+$ & $1790 \pm 35$
&
\begin{tabular}{c}
 $\sigma_N^E \rightarrow \pi \pi$   \\\hline  
 $\sigma_N^E \rightarrow KK$  \\\hline 
 $\sigma_N^E \rightarrow a_1(1260) \pi$ \\\hline
 $\sigma_N^E \rightarrow \eta \eta^{\prime}$ \\\hline
 $\sigma_N^E \rightarrow \eta \eta$ \\\hline
 $\sigma_N^E \rightarrow f_1(1285) \eta$ \\\hline
 $\sigma_N^E \rightarrow K_1 K$ \\\hline
 $\sigma_N^E \rightarrow \sigma_N \pi \pi$\\\hline
 Total
   \end{tabular}
&
\begin{tabular}{c}
 $\; \, 270 \pm 45$*   \\\hline
 $\; \, \; \; 70 \pm 40$*   \\\hline
 $\; \, 47 \pm \; \, 8$   \\\hline
 $\; \, 10 \pm \; \, 2$   \\\hline
 $\; \;  \, 7 \pm \, \, 1$   \\\hline
 
 $\; \; \; \, 1 \pm \; \, 0$   \\\hline
 $ \; \; 0 $   \\\hline
 $ \; \;  0 $ \\\hline
 $405 \pm 96$
   \end{tabular}
 
& 
\begin{tabular}{c}
Assigned to $f_0(1790)$;   \\
 mass, $\pi \pi$ and $KK$ \\decay widths 
 from Ref.\ \cite{BESII}. \\Other decays not (yet) measured.    
 
   \end{tabular}
\tabularnewline\hline

$\eta_{N}^{E}$ & $00^-$ & $1294 \pm 4$
&
$\eta_N^E \rightarrow \eta \pi \pi + \eta^{\prime} \pi \pi + \pi KK$
&
 $\; \; \; \, 7 \pm \; \, 3$   
& 
Assigned to $\eta(1295)$; PDG mass \cite{PDG}.  

 \tabularnewline\hline

$\eta_{S}^{E}$ & $00^-$ & $1432 \pm 10$
&
\begin{tabular}{c}
 $\eta_{S}^{E} \rightarrow K^{\star} K$   \\\hline  
 $\eta_{S}^{E} \rightarrow KK \pi$  \\\hline 
 $\eta_{S}^{E} \rightarrow \eta \pi \pi$ and $\eta^{\prime} \pi \pi$  \\\hline
 Total
   \end{tabular}
&
\begin{tabular}{c}
 $\; \, 128 ^{+204}_{-128}$   \\\hline
 $\; \; \; \, 28 ^{+41}_{-28} \; \,$   \\\hline
 suppressed   \\\hline
 $\; \, 156 ^{+245}_{-156}$
   \end{tabular}
   
& \begin{tabular}{c}
Assigned to $\eta(1440)$;   \\
  mass from Refs. \cite{BES1998,BES2000}.\\
  Full width $\sim$ 100 MeV at this mass \cite{BES2000}.\\
  $\Gamma_{\eta(1440) \rightarrow \eta \pi \pi}$ suppressed \cite{BES2000}.
 \end{tabular}

\tabularnewline\hline

$\sigma_{S}^{E}$ & $00^+$ & $1961 \pm 38$
&
\begin{tabular}{c}
 $\sigma_S^E \rightarrow KK$  \\\hline 
 $\sigma_S^E \rightarrow \eta \eta^{\prime}$ \\\hline
 $\sigma_S^E \rightarrow \eta \eta$ \\\hline
 $\sigma_S^E \rightarrow K_1 K$ \\\hline
 $\sigma_S^E \rightarrow \eta^{\prime} \eta^{\prime}$ \\\hline
 $\sigma_S^E \rightarrow \pi \pi$, $\rho \rho$ and $\omega \omega$   \\\hline 
 $\sigma_S^E \rightarrow a_1(1260) \pi$ and $f_1(1285) \eta$ \\\hline
 $\sigma_S^E \rightarrow \pi^E \pi$ and $\eta_N^E \eta$ \\\hline
 $\sigma_S^E \rightarrow \sigma_S \pi \pi$\\\hline
 Total
   \end{tabular}
&
\begin{tabular}{c}
  $\; \; \, 21 ^{+39}_{-21} \; \,$   \\\hline
  $\, 12 \pm  \,2$   \\\hline
  $\;\;\, 6 \pm  \, 1$   \\\hline
 $\; \; \, 2^{+5}_{-2}$ \\\hline
  $\; \, \, 1 \pm 0$   \\\hline
 suppressed   \\\hline
 suppressed   \\\hline
 suppressed \\\hline
 suppressed \\\hline
 $\; \; \; \, 42 ^{+47}_{-26} \; \,$
   \end{tabular}
   
& 
\begin{tabular}{c}

Candidate states: $f_0(2020)$;  \\
  $m_{f_0(2020)} = (1992 \pm 16)$ MeV and\\
  $\Gamma_{f_0(2020)} = (442 \pm 60)$ MeV \\\\and \\\\
  $f_0(2100)$;  \\
  $m_{f_0(2100)} = (2101 \pm 7)$ MeV and\\
  $\Gamma_{f_0(2101)} = 224^{+23}_{-21} $ MeV. \\\\
  Both require confirmation \cite{PDG}.
 \end{tabular}   
\tabularnewline\hline

$a_{0}^{E}$ & $10^+$ & $1790 \pm 35$
&
\begin{tabular}{c}
 $a_{0}^{E} \rightarrow \eta \pi$   \\\hline  
 $a_{0}^{E} \rightarrow KK$  \\\hline 
 $a_{0}^{E} \rightarrow \eta^{\prime} \pi$ \\\hline
 $a_{0}^{E} \rightarrow f_1(1285) \pi$ \\\hline
 $a_{0}^{E} \rightarrow a_1(1260) \eta$ \\\hline
 $a_{0}^{E} \rightarrow K_1 K$  \\\hline
 $a_{0}^{E} \rightarrow a_0(1450) \pi \pi$ \\\hline
 Total
   \end{tabular}
&
\begin{tabular}{c}
 $\; \; 72 \pm 12$   \\\hline
 $\; \; 70 \pm 40$   \\\hline
 $\; \, 32 \pm \; \, 5$   \\\hline
 $\; \, 16 \pm \; \, 3$   \\\hline
 $\; \; \; \, 1 \pm \; \, 0$   \\\hline
 $ \; \; 0 $ \\\hline
 $ \; \; 0 $ \\\hline
 $191 \pm 60$
   \end{tabular}
   
&
\begin{tabular}{c}

Candidate state: $a_0(1950)$;   \\
  $m_{a_0(1950)} = (1931 \pm 26)$ MeV and\\
  $\Gamma_{a_0(1950)} = (271 \pm 40)$ MeV \cite{BABAR}.\\\\ 
  Requires confirmation \cite{PDG}.
 \end{tabular}  
\tabularnewline\hline

$K_{0}^{\star E}$ & $\frac{1}{2}0^+$ & $1877 \pm 36$
&
\begin{tabular}{c}
  $K_{0}^{\star E}\rightarrow K \pi$  \\\hline 
 $K_{0}^{\star E} \rightarrow \eta^{\prime} K$   \\\hline  
 $K_{0}^{\star E} \rightarrow K_1 \pi$ \\\hline
 $K_{0}^{\star E} \rightarrow \eta K$ \\\hline
 $K_{0}^{\star E} \rightarrow a_1(1260) K$ \\\hline
 $K_{0}^{\star E} \rightarrow f_1(1285) K$  \\\hline
 $K_{0}^{\star E} \rightarrow K_1 \eta$  \\\hline
 $K_{0}^{\star E} \rightarrow K_0^{\star}(1430) \pi \pi$ \\\hline
 Total
   \end{tabular}
&
\begin{tabular}{c}
  $\; \; 51 \pm 35$   \\\hline
  $\;  24 \pm \,  4$   \\\hline
  $\; \;  \, 6 \pm \, 4$   \\\hline
 $\; \; \, 4^{+7}_{-4}$   \\\hline
 $\; \; \; \; 3 \pm \; \, 2$   \\\hline
 $\; \; \; \; 1 \pm \; \, 1$ \\\hline
 $ \; \; \, 0 $ \\\hline
 $ \; \; \, 0 $ \\\hline
 $\; \,  89 ^{+53}_{-50}$
   \end{tabular}
   
&  
\begin{tabular}{c}

Candidate state: $K_0^{\star}(1950)$;   \\
  $m_{K_0^{\star}(1950)} = (1945 \pm 22)$ MeV and\\
  $\Gamma_{K_0^{\star}(1950)} = (201 \pm 90)$ MeV. \\\\
  Requires confirmation \cite{PDG}.  
 \end{tabular}  

\tabularnewline\hline

$\pi^{E}$ & $10^-$ & $1294 \pm 4$
&
-
&
-   
& Width badly defined \\

& & & & & due to large errors of \\ & & & & & the experimental input data. 
\tabularnewline\hline

$K^{E}$ & $\frac{1}{2}0^-$ & $1366 \pm 6$
&
-
&
-   
& Width badly defined \\

& & & & & due to large errors of \\ & & & & & the experimental input data. 
\tabularnewline\hline

\end{tabular}
\caption{Decays and masses of the excited $\bar{q}q$ states. Widths marked as ``suppressed''
depend only on large-$N_c$ suppressed parameters that have been set to
zero. Widths marked with an asterisk are used as input; the others are predictions.}
\label{Decayst1}%
\end{table}
$\,$
\\
The consequences of $f_0(1790)$ input data are then as follows:

\begin{itemize}
 \item The excited states are generally rather narrow with the exception of $f_0(1790)$ and $\eta(1440)$
 whose full decay
 widths, considering the errors, are respectively between $\sim$ 300 MeV and $\sim$ 500 MeV and up to
 $\sim$ 400 MeV. The result for $f_0(1790)$ is congruent with the data
 published by LHCb \cite{LHCb}; the large interval for the $\eta(1440)$ width is a consequence of 
 parameter uncertainties, induced by ambiguities in the experimental input data.
 
 \item The excited pion and kaon states are also very susceptible to parameter uncertainties
 that lead to extremely large errors for the $\pi^E$ and $K^E$ decay widths [$\cal O$(1 GeV)]. 
 A definitive statement
 on these states is therefore not possible. Contrarily, in the case of $\eta(1295)$, the three decay widths accessible
 to our model (for $\eta_N^E \rightarrow \eta \pi \pi + \eta^{\prime} \pi \pi + \pi KK$) amount
 to $(7 \pm3)$ MeV and hence contribute very little to the overall decay width $\Gamma_{\eta(1295)}^{\text{total}} = (55 \pm 5)$ MeV.
 
 \item Analogously to the above point, parameter uncertainties also lead to extremely large
 width intervals for the decays of scalars into vectors. These decays are therefore omitted from Table
 \ref{Decayst1}, except for the large-$N_c$ suppressed decays $\sigma_S^E \rightarrow \rho \rho$
 and $\sigma_S^E \rightarrow \omega \omega$.
 
 \item Notwithstanding the above two points, we are able to predict more than 35 decay widths for
 all states in our model except $\pi^E$ and $K^E$. The overall correspondence of the model states to the
 experimental (unconfirmed) ones is generally rather good, although we note that our scalar $\bar{s}s$
 state appears to be too narrow to fully accommodate either of the $f_0(2020)$ and $f_0(2100)$ states.
 The mass of our isotriplet state $a_0^E$ is also somewhat smaller than that of $a_0(1950)$ --
 we will come back to this point in Sec.\ \ref{Decays5}.
 
\end{itemize}

\subsubsection{Hypothesis: $\eta(1295)$ and $\eta(1440)$ are excited $\bar{q}q$ states} \label{Decays4}

As indicated above, results presented in Table \ref{Decayst1} do not allow us to make a definitive statement
on all excited pseudoscalars.
However, the situation changes if the parameters $h_2^{\ast}$ and $h_3^{\ast}$ are determined with 
the help of the $\eta(1295)$ and $\eta(1440)$ decay widths.
\\
Using $\Gamma_{\eta_N^E \rightarrow \eta \pi \pi + \eta^{\prime} \pi \pi + \pi KK} = (55 \pm 5)$ MeV \cite{PDG} 
and $\Gamma_{\eta(1440) \rightarrow K^{\star}K}
= 26 \pm 3$ MeV (from Ref.\ \cite{BES1998}; our estimate for the error) we obtain
\begin{align}
h_{2}^{\ast} = 70 \pm 2 \text{, } h_{3}^{\ast} = 35 \pm 3 \;. \label{h23*2}
\end{align}
The parameters (\ref{h23*2}) are strongly constrained and there is a very good correspondence of 
the pseudoscalar decays to the experimental data in this case (see Table \ref{Decayst2}).
Nonetheless, there is a drawback: all scalar states become unobservable due to very broad decays into vectors.
Thus comparison of Tables \ref{Decayst1} and \ref{Decayst2} suggests that there is tension
between the simultaneous interpretation of $\eta(1295)$, $\pi(1300)$, $\eta(1440)$ and $K(1460)$
as well as the scalars as excited $\bar{q}q$ states.
A possible theoretical reason is that pseudoscalars above 1 GeV may have non-$\bar{q} q$ admixture.
Indeed sigma-model studies in Refs.\ \cite{Napsuciale2004,F1,F2,Fariborz:2007ai,Fariborz2009,F4,F5,Zebarjad} have
concluded that excited pseudoscalars with masses between 1 GeV and 1.5 GeV represent a mixture of $\bar{q}q$ and
$\bar{q}\bar{q}qq$ structures. In addition, the flux-tube model of Ref.\ \cite{Faddeev:2003aw} and a mixing formalism
based on the Ward identity in Ref.\ \cite{Cheng2009} lead to the conclusion that the pseudoscalar channel around
1.4 GeV is influenced by a glueball contribution. Hence a more complete description of these states would require
implementation of mixing scenarios in this channel\footnote{A similar mixing scenario may (as a matter
of principle) also exist in the case of the scalars discussed here. However, the amount of theoretical studies is
significantly smaller here: for example, a glueball contribution to $f_0(1790)$ has been discussed in
Refs.\ \cite{Bugg:2006uk,Bicudo:2006sd} while -- just as in our study -- the same resonance was found to be
compatible with an excited $\bar{q}q$ state in Ref.\ \cite{Vijande:2005jd}.}.\\
Note, however, that results of Table \ref{Decayst2} depend on the assumption that the total decay width of
$\eta(1295)$ is saturated by the three decay channels accessible to our model ($\eta \pi \pi$, $\eta^{\prime} \pi \pi$
and $K \pi \pi$). The level of justification for this assumption is currently uncertain \cite{PDG}.
Consequently we will not explore this scenario further.
\\
\begin{table}[h]
\centering%
\begin{tabular}{|c|c|c|c|c|@{}c@{}|}\hline
Model state
& 
$IJ^P$
& 
Mass (MeV)
&
Decay
&
Width (MeV)
&
Note  
\tabularnewline\hline

$\eta_{N}^{E}$ & $00^-$ & $1294 \pm 4$
&
$\eta_N^E \rightarrow \eta \pi \pi + \eta^{\prime} \pi \pi + \pi KK$
&
 $\; \; \; 55 \pm \; \, 5$*   
& 

Assigned to $\eta(1295)$; PDG mass \cite{PDG}.  

 \tabularnewline\hline

$\eta_{S}^{E}$ & $00^-$ & $1432 \pm 10$
&
\begin{tabular}{c}
 $\eta_{S}^{E} \rightarrow K^{\star} K$   \\\hline  
 $\eta_{S}^{E} \rightarrow KK \pi$  \\\hline 
 $\eta_{S}^{E} \rightarrow \eta \pi \pi$ and $\eta^{\prime} \pi \pi$  \\\hline
 Total
   \end{tabular}
&
\begin{tabular}{c}
 $\; \; \; 26 \pm  \; \, 3$*   \\\hline
 $\; \; \; \, 3 \pm \; \, 0$   \\\hline
 suppressed   \\\hline
  $\; \, 29 \pm \; \, 3$
   \end{tabular}
   
& \begin{tabular}{c}
Assigned to $\eta(1440)$;   \\
  mass and $K^{\star} K$ width \\
 from Refs. \cite{BES1998,BES2000}.
\\
Our estimate for $\Delta\Gamma_{\eta(1440) \rightarrow K^{\star} K}$.
 \end{tabular}

\tabularnewline\hline

$\pi^{E}$ & $10^-$ & $1294 \pm 4$
&
\begin{tabular}{c}
 $\pi^E \rightarrow \rho \pi$  \\\hline 
 $\pi^E \rightarrow 3\pi$ \\\hline
 $\pi^E \rightarrow K K \pi$ \\\hline
 Total
   \end{tabular}
&
\begin{tabular}{c}
 $368 \pm 37$   \\\hline
 $204 \pm 15$   \\\hline
 $\; \; \; \, 2 \pm \;  \, 0$   \\\hline
 $574 \pm 52$
   \end{tabular}
   
&  \begin{tabular}{c}
Assigned to $\pi(1300)$;   \\
  degenerate in mass with $\eta(1295)$\\
  according to Eq.\ (\ref{etaNE1}). \\
 Compares well with \\$\Gamma_{\pi(1300)} = (200-600)$ MeV \cite{PDG}.  
 \end{tabular}

\tabularnewline\hline

$K^{E}$ & $\frac{1}{2}0^-$ & $1366 \pm 6$
&
\begin{tabular}{c}
 $K^{E} \rightarrow K^{\star} \pi$   \\\hline  
 $K^{E} \rightarrow K \pi \pi$  \\\hline 
 $K^{E} \rightarrow \rho K$ \\\hline
 $K^E \rightarrow \omega K$ \\\hline
 $K^{E} \rightarrow K \pi \eta$ \\\hline
 Total
   \end{tabular}
&
\begin{tabular}{c}
 $112 \pm 11$   \\\hline
 $\; \; \, 35 \pm \; \; \, 4$   \\\hline
 $\; \; \, 20 \pm \; \; \, 2$   \\\hline
 $\; \; \; \; \, 7 \pm \; \; \, 1$   \\\hline
 $ \; \; 0 $   \\\hline
 $174 \pm 18$
   \end{tabular}
   
&\begin{tabular}{c}
Assigned to $K(1460)$;   \\
  $m_{K(1460)} \sim 1460$ MeV;\\
  $\Gamma_{K(1460)} \sim 260$ MeV \cite{PDG}.  
 \end{tabular}
\tabularnewline\hline

All scalars & - & As in Table \ref{Masses}
&
See Appendix \ref{Appendix}
&\begin{tabular}{c}
Calculated via   \\
Eqs.\ (\ref{SPP}), (\ref{SVP}), (\ref{SVV}),\\
 (\ref{Gamma3}) and Eq.\ (\ref{DeltaO}).  
 \end{tabular}

&\begin{tabular}{c}
Unobservable due to  \\
 extremely large decays \\ into vectors [$\cal O$(1 GeV)].  
 \end{tabular} 
\tabularnewline\hline

\end{tabular}
\caption{Decays and masses for the case where $\eta(1295)$ and $\eta(1440)$ 
are enforced as excited $\bar{q}q$ states. Widths marked with an asterisk were used as input.
Pseudoscalar observables compare fine
with experiment but the scalars are unobservable due to extremely broad decays into
vector mesons.}%
\label{Decayst2}%
\end{table}

\subsubsection{Is $a_0(1950)$ of BABAR Collaboration an excited $\bar{q}q$ state?} \label{Decays5}

\begin{figure}[h]
\begin{center}%
\begin{tabular}
[c]{cc}%
\resizebox{90mm}{!}{\includegraphics{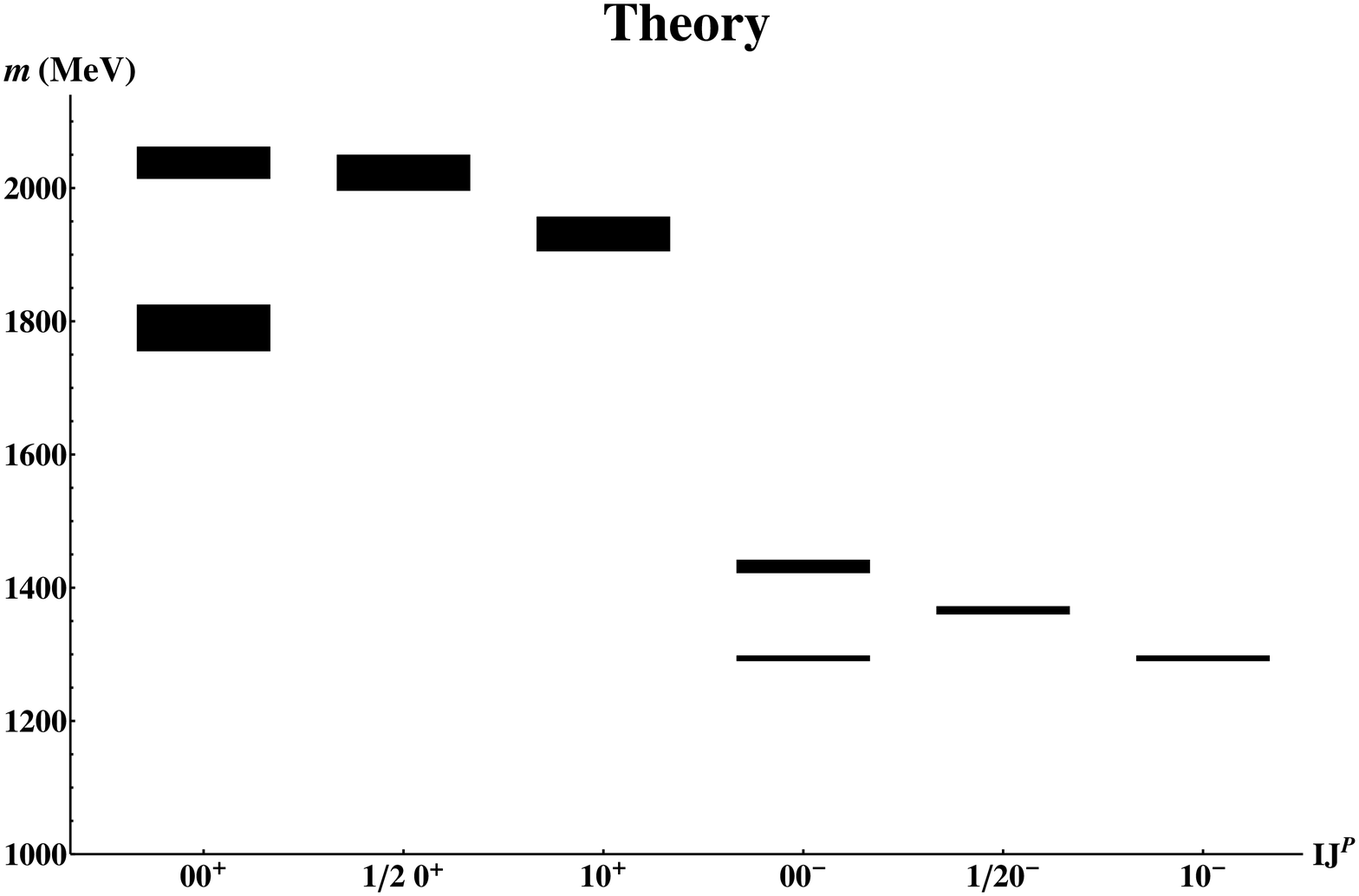}} &
\resizebox{90mm}{!}{\includegraphics{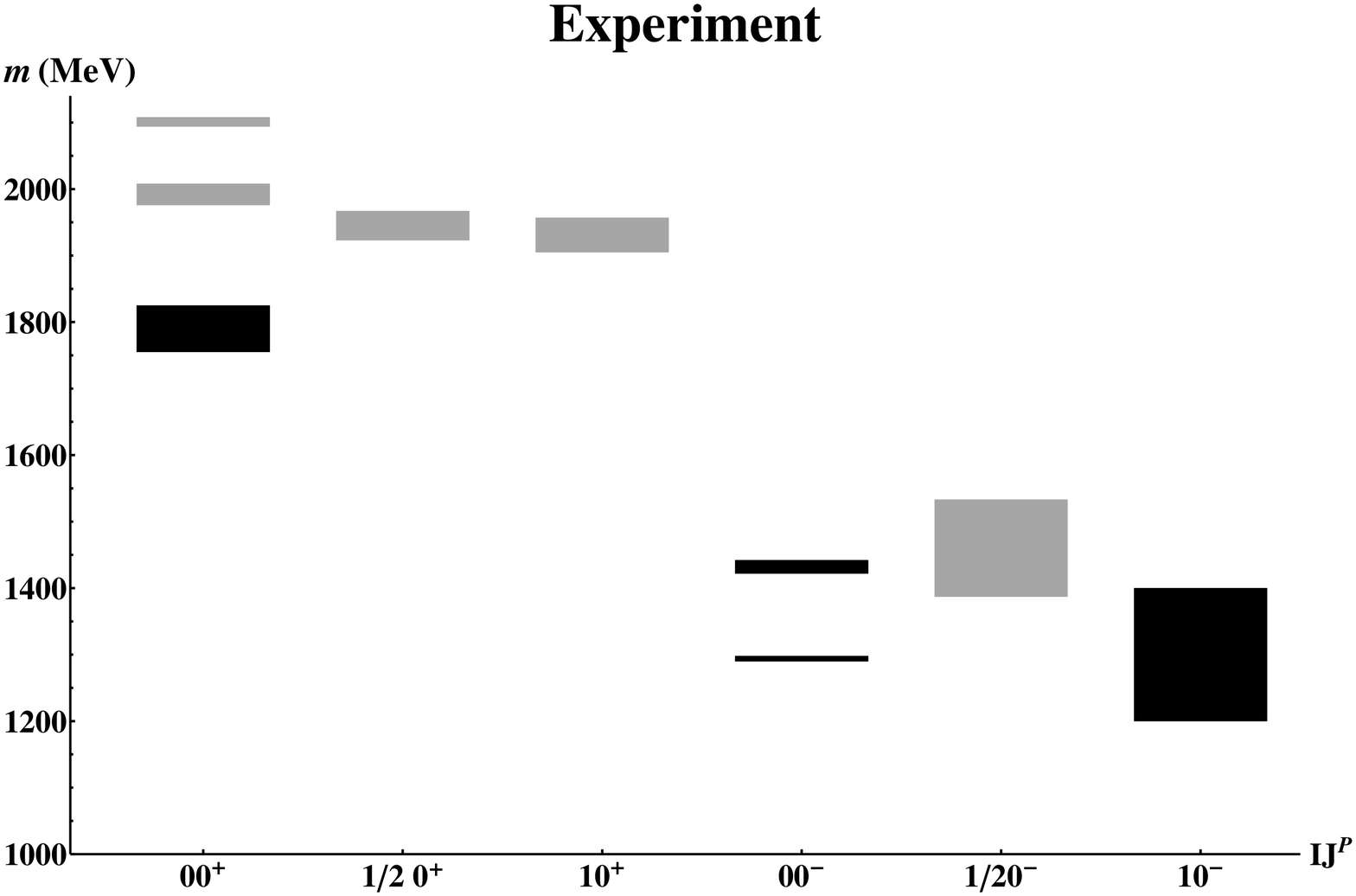}}
\end{tabular}
\end{center}
\caption{Masses of excited $\bar{q}q$ states with isospin $I$, total spin $J$ and parity $P$ from the Extended Linear Sigma Model (left) and masses from the experimental
data (right). Area thickness corresponds to mass uncertainties on both panels. The 
lower $00^+ (\equiv \sigma_N^E)$, both $00^-(\equiv \eta_N^E$ and $\eta_S^E)$ as well as the $10^+ (\equiv a_0^E)$ states from the left panel were used as input. Lightly shaded areas correspond to experimentally as yet unestablished states. Table \ref{Decayst3} contains the experimental assignment of the states on the left panel and a brief overview of their dynamics.}
\label{Figure1}%
\end{figure}

Encouraging results obtained in Sec.\ \ref{Decays3}, where $f_0(1790)$ was assumed to be an excited
$\bar{q}q$ state, can be used as a motivation to explore them further. As discussed in Sec.\ \ref{LagrangianS},
data analysis published recently by BABAR Collaboration has found evidence of an isotriplet state 
$a_0(1950)$ with mass $m_{a_0(1950)} = (1931 \pm 26)$ MeV and decay width $\Gamma_{a_0(1950)} = 
(271 \pm 40)$ MeV \cite{BABAR}.\\
Assuming that $f_0(1790)$ is an excited $\bar{q}q$ state (as already done in Sec. \ref{Decays3}), we can implement
$m_{a_0(1950)}$ obtained by BABAR as a large-$N_c$ suppressed effect in our model as follows. Mass terms for excited states
$\sigma_N^E$ and $\sigma_S^E$, Eqs.\ (\ref{sigmaNE1}) and (\ref{sigmaSE1}), can be modified by reintroduction
of the large-$N_c$ suppressed parameter $\kappa_2$ and now read
\begin{align}
m_{\sigma_{N}^{E}}^{2}  &  =C_{1}^{\ast}+\left(\frac{\xi_{2}}{2}+ 2 \kappa_2 \right)\phi_{N}%
^{2}\label{sigmaNE2}\\
m_{\sigma_{S}^{E}}^{2}  &  =C_{1}^{\ast}+2C_{2}^{\ast}+\left(\frac{\xi_{2}}{2}+ 2 \kappa_2 \right)
(\phi_{N}-2Z_{K}f_{K})^{2} \; .\label{sigmaSE2}
\end{align}
$\,$
\\
The other mass terms [Eqs.\ (\ref{a0E1}) - (\ref{etaSE1}), (\ref{KE1}) and (\ref{KSE1})] remain exactly the same;
$\kappa_2$ does not influence any decay widths.
We can now repeat the calculations described in Sec.\ \ref{Masses2} with the addition that the mass of 
our state $a_0^E$ corresponds exactly to that of $a_0(1950)$. We obtain
\begin{align}
C_{1}^{\ast }& =(2.4\pm 0.6)\cdot 10^{6}\text{ [MeV}^{2}\text{], }%
C_{2}^{\ast }=(2.5\pm 0.2)\cdot 10^{5}\text{ [MeV}^{2}\text{], }\xi
_{2}=57\pm 5\text{, } \kappa_2 = -10 \pm 3 \label{Massparameters2}\; .
\end{align}
Note that a non-vanishing value of $\kappa_2$ introduces mixing of $\sigma_N^E$ and $\sigma_S^E$ in
our Lagrangian (\ref{LagrangianE}). Its effect is, however, vanishingly small since the mixing angle is
$\sim 11^{\circ}$. 
\\
\\
Using the mass parameters (\ref{Massparameters2}) and the decay parameters (\ref{h23*}) we can repeat
the calculations of Sec.\ \ref{Decays3}.
Then our final results for the mass spectrum are presented and Fig. \ref{Figure1} and for the decays in Table \ref{Decayst3}.
The values of $m_{a_0^E}$, $m_{\sigma_S^E}$ and $m_{K_0^{\star E}}$ have changed in comparison to Table \ref{Decayst1} inducing an increased phase space. For this reason, the decay widths of the corresponding resonances have changed as well. All other results from 
Table \ref{Decayst1} have remained the same and are again included for clarity and convenience of the reader.

\begin{table}[th]
\centering%
\begin{tabular}{|c|c|c|c|c|@{}c@{}|}\hline
Model state
& 
$IJ^P$
& 
Mass (MeV)
&
Decay
&
Width (MeV)
&
Note  
\tabularnewline\hline

$\sigma_{N}^{E}$ & $00^+$ & $1790 \pm 35$*
&
\begin{tabular}{c}
$\sigma_N^E \rightarrow \pi \pi$   \\\hline  
 $\sigma_N^E \rightarrow KK$  \\\hline 
 $\sigma_N^E \rightarrow a_1(1260) \pi$ \\\hline
 $\sigma_N^E \rightarrow \eta \eta^{\prime}$ \\\hline
 $\sigma_N^E \rightarrow \eta \eta$ \\\hline
 $\sigma_N^E \rightarrow f_1(1285) \eta$ \\\hline
 $\sigma_N^E \rightarrow K_1 K$ \\\hline
 $\sigma_N^E \rightarrow \sigma_N \pi \pi$\\\hline
 Total
   \end{tabular}
&
\begin{tabular}{c}
 $\; \, 270 \pm 45$*   \\\hline
 $\; \, \; \; 70 \pm 40$*   \\\hline
 $\; \, 47 \pm \; \, 8$   \\\hline
 $\; \, 10 \pm \; \, 2$   \\\hline
 $\; \;  \, 7 \pm \, \, 1$   \\\hline

 $\; \; \; \, 1 \pm \; \, 0$   \\\hline
 $ \; \; 0 $   \\\hline
 $ \; \;  0 $ \\\hline
 $405 \pm 96$
   \end{tabular}
 
& 
\begin{tabular}{c}
Assigned to $f_0(1790)$;   \\
 mass, $\pi \pi$ and $KK$ \\decay widths 
 fixed to BES II data \cite{BESII}. \\Other decays not (yet) measured.    
 
   \end{tabular}
\tabularnewline\hline
$a_{0}^{E}$ & $10^+$ & $1931 \pm 26$*
&
\begin{tabular}{c}
 $a_{0}^{E} \rightarrow \eta \pi$   \\\hline  
 $a_{0}^{E} \rightarrow KK$  \\\hline 
 $a_{0}^{E} \rightarrow \eta^{\prime} \pi$ \\\hline
 $a_{0}^{E} \rightarrow f_1(1285) \pi$ \\\hline
 $a_{0}^{E} \rightarrow K_1 K$  \\\hline
 $a_{0}^{E} \rightarrow a_1(1260) \eta$ \\\hline
 $a_{0}^{E} \rightarrow a_0(1450) \pi \pi$ \\\hline
 Total
   \end{tabular}
&
\begin{tabular}{c}
 $\; \; 94 \pm 16$   \\\hline
 $\; \; 94 \pm 54$   \\\hline
 $\; \, 48 \pm \; \, 8$   \\\hline
 $\; \, 28 \pm \; \, 5$   \\\hline
 $\; \; \; \, 9 \pm \; \, 5$ \\\hline
 $\; \; \; \, 6 \pm \; \, 1$   \\\hline
 $\; \; \; \, 1 \pm \; \, 1$ \\\hline
 $280 \pm 90$
   \end{tabular}
   
&
\begin{tabular}{c}

Candidate state: $a_0(1950)$ \\recently measured by BABAR;   \\
  $m_{a_0(1950)} = (1931 \pm 26)$ MeV and\\
  $\Gamma_{a_0(1950)} = (271 \pm 40)$ MeV \cite{BABAR}.\\\\ 
  Requires confirmation \cite{PDG}.
 \end{tabular}  
\tabularnewline\hline

$\eta_{N}^{E}$ & $00^-$ & $ 1294 \pm\, \,4 $*
&
$\eta_N^E \rightarrow \eta \pi \pi + \eta^{\prime} \pi \pi + \pi KK$
&
 $\; \; \; \, 7 \pm \; \, 3$   
& 
Assigned to $\eta(1295)$; PDG mass \cite{PDG}.  

 \tabularnewline\hline

$\eta_{S}^{E}$ & $00^-$ & $1432 \pm 10$*
&
\begin{tabular}{c}
 $\eta_{S}^{E} \rightarrow K^{\star} K$   \\\hline  
 $\eta_{S}^{E} \rightarrow KK \pi$  \\\hline 
 $\eta_{S}^{E} \rightarrow \eta \pi \pi$ and $\eta^{\prime} \pi \pi$  \\\hline
 Total
   \end{tabular}
&
\begin{tabular}{c}
 $\; \, 128 ^{+204}_{-128}$   \\\hline
 $\; \; \; \, 28 ^{+41}_{-28} \; \,$   \\\hline
 suppressed   \\\hline
 $\; \, 156 ^{+245}_{-156}$
   \end{tabular}
   
& \begin{tabular}{c}
Assigned to $\eta(1440)$;   \\
  mass from BES data \cite{BES1998,BES2000}.\\
  Full width $\sim$ 100 MeV at this mass \cite{BES2000}.\\
  $\Gamma_{\eta(1440) \rightarrow \eta \pi \pi}$ suppressed \cite{BES2000}.
 \end{tabular}

\tabularnewline\hline

$\sigma_{S}^{E}$ & $00^+$ & $2038 \pm 24 \;$
&
\begin{tabular}{c}
 $\sigma_S^E \rightarrow KK$  \\\hline 
 $\sigma_S^E \rightarrow \eta \eta^{\prime}$ \\\hline
 $\sigma_S^E \rightarrow \eta \eta$ \\\hline
 $\sigma_S^E \rightarrow K_1 K$ \\\hline
 $\sigma_S^E \rightarrow \eta^{\prime} \eta^{\prime}$ \\\hline
 $\sigma_S^E \rightarrow \pi \pi$, $\rho \rho$ and $\omega \omega$    \\\hline 
 $\sigma_S^E \rightarrow a_1(1260) \pi$ and $f_1(1285) \eta$ \\\hline
 $\sigma_S^E \rightarrow \pi^E \pi$ and $\eta_N^E \eta$ \\\hline
 $\sigma_S^E \rightarrow \sigma_S \pi \pi$\\\hline
 Total
   \end{tabular}
&
\begin{tabular}{c}
$\; \; \; \, 24 ^{+46}_{-24} \; \,$   \\\hline
 $\, \, 16 \pm \, 3$   \\\hline
 $\; \, \, 7 \pm 1$   \\\hline
 $\; \, \, 4^{+8}_{-4}$ \\\hline
 $\; \, \, 1 \pm 0$   \\\hline
 suppressed   \\\hline
 suppressed   \\\hline
 suppressed \\\hline
 suppressed \\\hline
 $\; \; \; \, 52 ^{+58}_{-32} \; \,$
   \end{tabular}
   
& 
\begin{tabular}{c}

Candidate states: $f_0(2020)$;  \\
  $m_{f_0(2020)} = (1992 \pm 16)$ MeV and\\
  $\Gamma_{f_0(2020)} = (442 \pm 60)$ MeV \\\\and \\\\
  $f_0(2100)$;  \\
  $m_{f_0(2100)} = (2101 \pm 7)$ MeV and\\
  $\Gamma_{f_0(2101)} = 224^{+23}_{-21} $ MeV. \\\\
  Both require confirmation \cite{PDG}.
 \end{tabular}   
\tabularnewline\hline

$K_{0}^{\star E}$ & $\frac{1}{2}0^+$ & $2023 \pm 27\;$
&
\begin{tabular}{c}
 $K_{0}^{\star E} \rightarrow \eta^{\prime} K$   \\\hline  
 $K_{0}^{\star E}\rightarrow K \pi$  \\\hline 
 $K_{0}^{\star E} \rightarrow K_1 \pi$ \\\hline
 $K_{0}^{\star E} \rightarrow a_1(1260) K$ \\\hline
 $K_{0}^{\star E} \rightarrow \eta K$ \\\hline
 $K_{0}^{\star E} \rightarrow f_1(1285) K$  \\\hline
 $K_{0}^{\star E} \rightarrow K_1 \eta$  \\\hline
 $K_{0}^{\star E} \rightarrow K_0^{\star}(1430) \pi \pi$ \\\hline
 Total
   \end{tabular}
&
\begin{tabular}{c}
 $\; \; 72 \pm  12$   \\\hline
 $\; \; 66 \pm 46$   \\\hline
 $\; \; 10 \pm \; \, 7$   \\\hline
 $\; \; \; 6 \pm \, 4$   \\\hline
 $\; \; \; 6^{+9}_{-6}$   \\\hline
 $\; \; \; \, 2 \pm \; 1$ \\\hline
 $ \; \; 0 $ \\\hline
 $ \; \; 0 $ \\\hline
 $\,  162 ^{+79}_{-76}$
   \end{tabular}
   
&  
\begin{tabular}{c}

Candidate state: $K_0^{\star}(1950)$;   \\
  $m_{K_0^{\star}(1950)} = (1945 \pm 22)$ MeV and\\
  $\Gamma_{K_0^{\star}(1950)} = (201 \pm 90)$ MeV. \\\\
  Requires confirmation \cite{PDG}.  
 \end{tabular}  

\tabularnewline\hline

$\pi^{E}$ & $10^-$ & $1294 \pm 4 \;$
&
-
&
-   
& Width badly defined \\

& & & & & due to large errors of \\ & & & & & the experimental input data. 
\tabularnewline\hline

$K^{E}$ & $\frac{1}{2}0^-$ & $1366 \pm 6\;$
&
-
&
-   
& Width badly defined \\

& & & & & due to large errors of \\ & & & & & the experimental input data. 
\tabularnewline\hline

\end{tabular}
\caption{Final results:\ decays and masses of the excited $\bar{q}q$ states. Widths marked as ``suppressed''
depend only on large-$N_c$ suppressed parameters that have been set to
zero. Masses/widths marked with (*) are used as input; the others are predictions.}
\label{Decayst3}%
\end{table}
$\,$
\\
The consequences are as follows:

\begin{itemize}
 \item The decay width of $a_0^E$ is now $\Gamma_{a_0^E} = (280 \pm 90)$ MeV;
 it overlaps fully with $\Gamma_{a_0(1950)} = (271 \pm 40)$ MeV measured by BABAR. Hence,
 if $a_0(1950)$ is confirmed in future measurements, it will represent a very good candidate for the excited 
 isotriplet $\bar{n}n$ state.
 
 \item The mass of $\sigma_S^E$ is between those of $f_0(2020)$ and $f_0(2100)$. 
 Judging by the quantum numbers, either of these resonances
 could represent a (predominant) $\bar{s}s$ state; an option is also that the excited $\bar{s}s$ state
 with $IJ^{PC} = 00^{++}$ has not yet been observed in this energy region.
 However, one must also remember the possibility that
 $\bar{q}q$-glueball mixing (neglected here) may change masses as well as decay patterns.
 The decay width of $\sigma_S^E$ is rather narrow (up to 110 MeV) but this may change if mixing
 effects happen to be large.
 
 \item The mass of $K_{0}^{\star E}$ is qualitatively (within $\sim$ 100 MeV) congruent with that of
 $K_{0}^{\star} (1950)$; the widths overlap within 1 $\sigma$. Hence,
 if $K_{0}^{\star} (1950)$ is confirmed in future measurements, it will represent a very good candidate for the
 excited scalar kaon.
 
 \item Conclusions for all other states remain as in Sec.\ \ref{Decays3}.
\end{itemize}

\section{Conclusion} \label{Conclusion}

We have studied masses and decays of excited scalar and pseudoscalar $\bar{q}q$ states ($q = u,d,s$ quarks) 
in the Extended Linear Sigma Model (eLSM) that, in addition, contains ground-state scalar, pseudoscalar, 
vector and axial-vector mesons.\\
Our main objective was to study the assumption that the $f_0(1790)$ resonance
is an excited $\bar{n}n$ state. This assignment was motivated by the observation in BES \cite{BESII}
and LHCb \cite{LHCb} data that
the resonance couples mostly to pions and by the theoretical statement that the $\bar{n}n$ ground state is 
contained in the physical spectrum below $f_0(1790)$. Furthermore, the assumption was also
tested that the $a_0(1950)$ resonance, whose discovery was recently claimed by the BABAR Collaboration
\cite{BABAR}, represents the isotriplet partner of $f_0(1790)$.\\
Using the mass, $2\pi$ and $2K$ decay widths of $f_0(1790)$, the mass of $a_0(1950)$ and the masses
of the pseudoscalar isosinglets $\eta(1295)$ and $\eta(1440)$ our model predicts more than 35 decays
for all excited states except for the excited pion and kaon (where extremely large uncertainties are present
due to experimental ambiguities). All numbers are collected in Table \ref{Decayst3}.\\
In essence: the $f_0(1790)$ resonance emerges as the broadest excited $\bar{q}q$ state in the scalar channel with
$\Gamma_{f_0(1790)} = (405 \pm 96)$ MeV; $a_0(1950)$, if confirmed, represents a very good candidate
for the excited $\bar{q}q$ state; $K_0^{\star}(1950)$, if confirmed, represents a very good candidate for the excited scalar kaon.\\
Our excited isoscalar $\bar{s}s$ state has a mass of $(2038 \pm 24)$ MeV, placed between the masses
of the nearby $f_0(2020)$ and $f_0(2100)$ resonances; also, its width is relatively small 
($\leq 110$ MeV). We conclude that, although any of these resonances may in principle represent 
a $\bar{q}q$ state,
the introduction of mixing effects (particularly with a glueball state) may be necessary to further 
elucidate their structure.\\
Our results also imply a quite small contribution of the $\eta \pi \pi$, $\eta^{\prime} \pi \pi$
and $\pi KK$ decays to the overall width of $\eta(1295)$. For $\eta(1440)$, the decay width is 
compatible with any value up to $\sim 400$ MeV (ambiguities due to uncertainty in experimental
input data). \\
It is also possible to implement $\Gamma_{\eta(1295)}^{\text{total}}
\equiv \Gamma_{\eta(1295) \rightarrow \eta \pi \pi + \eta^{\prime} \pi \pi + \pi KK}$ 
and  $\Gamma_{\eta(1440) \rightarrow K^{\star}K}$ exactly as in the data of 
PDG \cite{PDG} and BES \cite{BES1998}. Then $\pi(1300)$ and $K(1460)$ are
quite well described as excited $\bar{q}q$ states -- but the scalars are unobservably
broad (see Table \ref{Decayst2}). Hence, in this case, there appears to be tension 
between the simultaneous description of $\eta(1295)$, $\pi(1300)$, $\eta(1440)$ 
and $K(1460)$ and their scalar counterparts as excited $\bar{q}q$ states. This scenario is,
however, marred by experimental uncertainties: for example, it is not at all clear 
if the width of $\eta(1295)$ is indeed saturated by the $\eta \pi \pi$, 
$\eta^{\prime} \pi \pi$ and $\pi KK$ decays. It could therefore only be explored further
when (very much needed) new experimental data arrive -- from BABAR, BES, LHCb or PANDA
\cite{PANDA} and NICA \cite{NICA}.

\section*{Acknowledgments}
We are grateful to D.~Bugg, C.~Fischer and A.~Rebhan for extensive discussions. The collaboration with Stephan 
H\"{u}bsch within a Project Work at TU Wien is also gratefully acknowledged. 
The work of D.~P.\ is supported by the Austrian 
Science Fund FWF, project no.\ P26366. The work of F.~G.\ 
is supported by the Polish National Science Centre NCN through the OPUS project
nr.\ 2015/17/B/ST2/01625.

\appendix

\section{Interaction Lagrangians}

\label{Appendix}

Here we collect all interaction Lagrangians that are used for calculations of
decay widths throughout this article. Vertices for large-$N_{c}$ suppressed
decays are not included but briefly discussed after each Lagrangian in which they appear.

\subsection{Lagrangian for $\sigma_{N}^{E}$}

The Lagrangian reads:%

\begin{align}
\mathcal{L}_{\sigma_{N}^{E}}  &  =\,\frac{1}{2}(h_{2}^{\star}-h_{3}^{\star
})w_{a_{1}}^{2}Z_{\pi}^{2}\phi_{N}\,\sigma_{N}^{E}\left[  (\partial_{\mu}%
\eta_{N})^{2}+(\partial_{\mu}\vec{\pi})^{2}\right]  +\frac{1}{2}\left(
h_{2}^{\star}\phi_{N}-\sqrt{2}h_{3}^{\star}\phi_{S}\right)  w_{K_{1}}^{2}%
Z_{K}^{2}\,\sigma_{N}^{E}\left(  \partial_{\mu}\bar{K}^{0}\partial^{\mu}%
K^{0}+\partial_{\mu}K^{-}\partial^{\mu}K^{+}\right) \nonumber\\
&  +(h_{2}^{\star}-h_{3}^{\star})w_{a_{1}}Z_{\pi}\phi_{N}\,\sigma_{N}%
^{E}\left(  f_{1N}^{\mu}\partial_{\mu}\eta_{N}+\vec{a_{1}}^{\mu}\cdot
\partial_{\mu}\vec{\pi}\right) \nonumber\\
&  +\frac{1}{2}\left(  h_{2}^{\star}\phi_{N}-\sqrt{2}h_{3}^{\star}\phi
_{S}\right)  w_{K_{1}}Z_{K}\,\sigma_{N}^{E}\left(  \bar{K}_{1\mu}^{0}%
\partial^{\mu}K^{0}+K_{1\mu}^{-}\partial^{\mu}K^{+}\text{ + h.c.}\right)
\nonumber\\
&  +\frac{1}{2}(h_{2}^{\star}+h_{3}^{\star})\phi_{N}\,\sigma_{N}^{E}\left[
(\omega_{N}^{\mu})^{2}+(\vec{\rho}^{\mu})^{2}\right]  +\frac{1}{2}\left(
h_{2}^{\star}\phi_{N}+\sqrt{2}h_{3}^{\star}\phi_{S}\right)  \,\sigma_{N}%
^{E}\left(  \bar{K}_{\mu}^{\star0}K^{\star\mu0}+K_{\mu}^{\star-}K^{\star\mu
+}\right) \nonumber\\
&  -\xi_{2}Z_{\pi}\phi_{N}\,\sigma_{N}^{E}\vec{\pi}^{E}\cdot\vec{\pi}%
-g_{1}^{E}w_{a_{1}}Z_{\pi}\,\sigma_{N}^{E}\partial_{\mu}\vec{\pi}^{E}%
\cdot\partial^{\mu}\vec{\pi} +\frac{1}{2}(h_{2}^{\star}-h_{3}^{\star}%
)w_{a_{1}}^{2}Z_{\pi}^{2}\,\sigma_{N}^{E}\sigma_{N}(\partial_{\mu}\vec{\pi
})^{2}\;.
\end{align}
$\,$
\\
Note: the decay $\sigma_{N}^{E}\rightarrow\eta_{S}\eta_{S}$ ($\sim\kappa_{1}$,
$h_{1}^{\star}$) is large-$N_{c}$ suppressed.

\subsection{Lagrangian for $\sigma_{S}^{E}$}

The Lagrangian reads:%

\begin{align}
\mathcal{L}_{\sigma_{S}^{E}}  &  =\,(h_{2}^{\star}-h_{3}^{\star})w_{f_{1S}%
}^{2}Z_{\eta_{S}}^{2}\phi_{S}\,\sigma_{S}^{E}(\partial_{\mu}\eta_{S})^{2}
+\left(  h_{2}^{\star}\phi_{S}-\frac{h_{3}^{\star}}{\sqrt{2}}\phi_{N}\right)
w_{K_{1}}^{2}Z_{K}^{2}\,\sigma_{S}^{E}\left(  \partial_{\mu}\bar{K}%
^{0}\partial^{\mu}K^{0}+\partial_{\mu}K^{-}\partial^{\mu}K^{+}\right)
\nonumber\\
&  +\left(  h_{2}^{\star}\phi_{S}-\frac{h_{3}^{\star}}{\sqrt{2}}\phi
_{N}\right)  w_{K_{1}}Z_{K}\,\sigma_{S}^{E}\left(  \bar{K}_{1\mu}^{0}%
\partial^{\mu}K^{0}+K_{1\mu}^{-}\partial^{\mu}K^{+}\text{ + h.c.}\right)
\nonumber\\
&  +\left(  h_{2}^{\star}\phi_{S}+\frac{h_{3}^{\star}}{\sqrt{2}}\phi
_{N}\right)  \,\sigma_{S}^{E}\left(  \bar{K}_{\mu}^{\star0}K^{\star\mu
0}+K_{\mu}^{\star-}K^{\star\mu+}\right)  \;.
\end{align}
$\,$
\\
Note: the decays $\sigma_{S}^{E}\rightarrow\pi\pi$ ($\sim\kappa_{1}$,
$h_{1}^{\star}$), $\sigma_{S}^{E}\rightarrow\eta_{N}\eta_{N}$ ($\sim\kappa
_{1}$, $h_{1}^{\star}$), $\sigma_{S}^{E}\rightarrow\rho\rho$ ($\sim
h_{1}^{\star}$), $\sigma_{S}^{E}\rightarrow\omega_{N}\omega_{N}$ ($\sim
h_{1}^{\star}$), $\sigma_{S}^{E}\rightarrow a_{1}\pi$ ($\sim h_{1}^{\star}$),
$\sigma_{S}^{E}\rightarrow f_{1N}\eta_{N}$ ($\sim h_{1}^{\star}$), $\sigma
_{S}^{E}\rightarrow\pi^{E}\pi$ ($\sim\kappa_{2}$), $\sigma_{S}^{E}%
\rightarrow\eta_{N}^{E}\eta_{N}$ ($\sim\kappa_{2}$) and $\sigma_{S}%
^{E}\rightarrow\sigma_{S}\pi\pi$ ($\sim\kappa_{1}$, $h_{1}^{\star}$) are
large-$N_{c}$ suppressed.

\subsection{Lagrangian for $a_{0}^{E}$}

The Lagrangian reads (only $a_{0}^{0E}$ included; decays of $a_{0}^{\pm E}$
follow from isospin symmetry):%

\begin{align}
\mathcal{L}_{a_{0}^{E}}  &  =\,(h_{2}^{\star}-h_{3}^{\star})w_{a_{1}}%
^{2}Z_{\pi}^{2}\phi_{N}\,a_{0}^{0E}\partial_{\mu}\pi^{0}\partial^{\mu}\eta_{N}
-\frac{1}{2}\left(  h_{2}^{\star}\phi_{N}-\sqrt{2}h_{3}^{\star}\phi
_{S}\right)  w_{K_{1}}^{2}Z_{K}^{2}\,a_{0}^{0E}\left(  \partial_{\mu}\bar
{K}^{0}\partial^{\mu}K^{0}-\partial_{\mu}K^{-}\partial^{\mu}K^{+}\right)
\nonumber\\
&  +(h_{2}^{\star}-h_{3}^{\star})w_{a_{1}}Z_{\pi}\phi_{N}\,a_{0}^{0E}\left(
f_{1N}^{\mu}\partial_{\mu}\pi^{0}+a_{1}^{\mu0}\partial_{\mu}\eta_{N}\right)
\nonumber\\
&  -\frac{1}{2}\left(  h_{2}^{\star}\phi_{N}-\sqrt{2}h_{3}^{\star}\phi
_{S}\right)  w_{K_{1}}Z_{K}\,a_{0}^{0E}\left(  \bar{K}_{1\mu}^{0}\partial
^{\mu}K^{0}-K_{1\mu}^{-}\partial^{\mu}K^{+}\text{ + h.c.}\right) \nonumber\\
&  +(h_{2}^{\star}+h_{3}^{\star})\phi_{N}\,a_{0}^{0E}\rho_{\mu}^{0}\omega
_{N}^{\mu}-\frac{1}{2}\left(  h_{2}^{\star}\phi_{N}+\sqrt{2}h_{3}^{\star}%
\phi_{S}\right)  \,a_{0}^{0E}\left(  \bar{K}_{\mu}^{\star0}K^{\star\mu
0}-K_{\mu}^{\star-}K^{\star\mu+}\right) \nonumber\\
&  -\xi_{2}Z_{\pi}\phi_{N}\,a_{0}^{0E}\eta_{N}^{E}\pi^{0}-g_{1}^{E}w_{a_{1}%
}Z_{\pi}\,a_{0}^{0E}\partial_{\mu}\eta_{N}^{E}\partial^{\mu}\pi^{0}\nonumber\\
&  +\frac{1}{2}(h_{2}^{\star}+h_{3}^{\star})w_{a_{1}}^{2}Z_{\pi}^{2}%
\,a_{0}^{0E}a_{0}^{0}(\partial_{\mu}\vec{\pi})^{2}-h_{3}^{\star}w_{a_{1}}%
^{2}Z_{\pi}^{2}\,a_{0}^{0E}\partial_{\mu}\pi^{0}\left(  \vec{a}_{0}%
\cdot\partial^{\mu}\vec{\pi}\right)  \;.
\end{align}

\subsection{Lagrangian for $K_{0}^{\star E}$}

The Lagrangian reads (only $K_{0}^{\star0E}$ included; decays of other
$K_{0}^{\star E}$ components follow from isospin symmetry):%

\begin{align}
\mathcal{L}_{K_{0}^{\star E}}  &  =\frac{1}{4}\left[  h_{2}^{\star}\left(
\phi_{N}+\sqrt{2}\phi_{S}\right)  -2h_{3}^{\star}\phi_{N}\right]  w_{a_{1}%
}w_{K_{1}}Z_{\pi}Z_{K}\,K_{0}^{\star0E}\left(  \partial_{\mu}\bar{K}%
^{0}\partial^{\mu}\eta_{N}-\partial_{\mu}\bar{K}^{0}\partial^{\mu}\pi
^{0}+\sqrt{2}\partial_{\mu}K^{-}\partial^{\mu}\pi^{+}\right) \nonumber\\
&  +\frac{1}{2\sqrt{2}}\left[  h_{2}^{\star}\left(  \phi_{N}+\sqrt{2}\phi
_{S}\right)  -2\sqrt{2}h_{3}^{\star}\phi_{S}\right]  w_{f_{1S}}w_{K_{1}%
}Z_{\eta_{S}}Z_{K}\,K_{0}^{\star0E}\partial_{\mu}\bar{K}^{0}\partial^{\mu}%
\eta_{S}\nonumber\\
&  +\frac{1}{4}\left[  h_{2}^{\star}\left(  \phi_{N}+\sqrt{2}\phi_{S}\right)
-2h_{3}^{\star}\phi_{N}\right]  w_{K_{1}}Z_{K}\,K_{0}^{\star0E}\left(
f_{1N}^{\mu}\partial_{\mu}\bar{K}^{0}-a_{1}^{\mu0}\partial_{\mu}\bar{K}%
^{0}+\sqrt{2}a_{1}^{\mu+}\partial_{\mu}K^{-}\right) \nonumber\\
&  +\frac{1}{4}\left[  h_{2}^{\star}\left(  \phi_{N}+\sqrt{2}\phi_{S}\right)
-2h_{3}^{\star}\phi_{N}\right]  w_{a_{1}}Z_{\pi}\,K_{0}^{\star0E}\left(
\bar{K}_{1\mu}^{0}\partial^{\mu}\eta_{N}-\bar{K}_{1\mu}^{0}\partial^{\mu}%
\pi^{0}+\sqrt{2}K_{1\mu}^{-}\partial^{\mu}\pi^{+}\right) \nonumber\\
&  +\frac{1}{2\sqrt{2}}\left[  h_{2}^{\star}\left(  \phi_{N}+\sqrt{2}\phi
_{S}\right)  -2\sqrt{2}h_{3}^{\star}\phi_{S}\right]  w_{f_{1S}}Z_{\eta_{S}%
}\,K_{0}^{\star0E}\bar{K}_{1\mu}^{0}\partial^{\mu}\eta_{S}\nonumber\\
&  +\frac{1}{4}\left[  h_{2}^{\star}\left(  \phi_{N}+\sqrt{2}\phi_{S}\right)
+2h_{3}^{\star}\phi_{N}\right]  \,K_{0}^{\star0E}\left(  \bar{K}_{\mu}%
^{\star0}\omega_{N}^{\mu}-\bar{K}_{\mu}^{\star0}\rho^{\mu0}+\sqrt{2}K_{\mu
}^{\star-}\rho^{\mu+}\right) \nonumber\\
&  -\frac{1}{4}\left[  2\xi_{2}\phi_{N}-\lambda_{2}^{\star}\left(  \phi
_{N}-\sqrt{2}\phi_{S}\right)  \right]  Z_{K}\,K_{0}^{\star0E}\left(  \bar
{K}^{0}\eta_{N}^{E}-\bar{K}^{0}\pi^{0E}+\sqrt{2}K^{-}\pi^{+E}\right)
\nonumber\\
&  -\frac{1}{2}g_{1}^{E}w_{K_{1}}Z_{K}\,K_{0}^{\star0E}\left(  \partial_{\mu
}\bar{K}^{0}\partial^{\mu}\eta_{N}^{E}-\partial_{\mu}\bar{K}^{0}\partial^{\mu
}\pi^{0E}+\sqrt{2}\partial_{\mu}K^{-}\partial^{\mu}\pi^{+E}\right) \nonumber\\
&  +\frac{1}{2}g_{1}^{E}w_{K_{1}}Z_{K}\,\partial_{\mu}K_{0}^{\star0E}\left(
\partial^{\mu}\bar{K}^{0}\eta_{N}^{E}-\partial^{\mu}\bar{K}^{0}\pi^{0E}%
+\sqrt{2}\partial^{\mu}K^{-}\pi^{+E}\right) \nonumber\\
&  +\frac{1}{\sqrt{2}}\xi_{2}Z_{\pi}\phi_{S}\,K_{0}^{\star0E}\left(  \bar
{K}^{0E}\pi^{0}-\sqrt{2}K^{-E}\pi^{+}\right)  +\frac{1}{2}g_{1}^{E}w_{a_{1}%
}Z_{\pi}\,K_{0}^{\star0E}\left(  \partial_{\mu}\bar{K}^{0E}\partial^{\mu}%
\pi^{0}-\sqrt{2}\partial_{\mu}K^{-E}\partial^{\mu}\pi^{+}\right) \nonumber\\
&  -\frac{1}{2}g_{1}^{E}w_{a_{1}}Z_{\pi}\,\partial_{\mu}K_{0}^{\star0E}\left(
\bar{K}^{0E}\partial^{\mu}\pi^{0}-\sqrt{2}K^{-E}\partial^{\mu}\pi^{+}\right)
\nonumber\\
&  +\frac{1}{4}h_{2}^{\star}w_{a_{1}}^{2}Z_{\pi}^{2}Z_{K_{S}}\,K_{0}^{\star
0E}\bar{K}_{0}^{\star0}(\partial_{\mu}\vec{\pi})^{2}+\frac{i}{4}(h_{2}^{\star
}-2h_{3}^{\star})w_{a_{1}}w_{K^{\star}}^{\ast}Z_{\pi}^{2}Z_{K_{S}}%
\,K_{0}^{\star0E}\pi^{0}\partial_{\mu}\bar{K}_{0}^{\star0}\partial^{\mu}%
\pi^{0}\nonumber\\
&  -ih_{3}^{\star}w_{a_{1}}w_{K^{\star}}^{\ast}Z_{\pi}^{2}Z_{K_{S}}%
\,K_{0}^{\star0E}\pi^{-}\partial_{\mu}\bar{K}_{0}^{\star0}\partial^{\mu}%
\pi^{+}+\frac{i}{2}h_{2}^{\star}w_{a_{1}}w_{K^{\star}}^{\ast}Z_{\pi}%
^{2}Z_{K_{S}}\,K_{0}^{\star0E}\pi^{+}\partial_{\mu}\bar{K}_{0}^{\star
0}\partial^{\mu}\pi^{-}\nonumber\\
&  +\frac{i}{2\sqrt{2}}(h_{2}^{\star}+2h_{3}^{\star})w_{a_{1}}w_{K^{\star}%
}^{\ast}Z_{\pi}^{2}Z_{K_{S}}\,K_{0}^{\star0E}\left(  \pi^{+}\partial_{\mu
}K_{0}^{\star-}\partial^{\mu}\pi^{0}-\pi^{0}\partial_{\mu}K_{0}^{\star
-}\partial^{\mu}\pi^{+}\right)  \;.
\end{align}

\subsection{Lagrangian for $\eta_{N}^{E}$}

Only three-body decays into pseudoscalars are kinematically allowed for this particle:%

\begin{align}
\mathcal{L}_{\eta_{N}^{E}}  &  =\,\frac{1}{2}(h_{2}^{\star}-h_{3}^{\star
})w_{a_{1}}^{2}Z_{\pi}^{3}\,\eta_{N}^{E}\eta_{N}(\partial_{\mu}\vec{\pi}%
)^{2}+(h_{2}^{\star}-h_{3}^{\star})w_{a_{1}}^{2}Z_{\pi}^{3}\,\eta_{N}%
^{E}\left(  \partial_{\mu}\eta_{N}\partial^{\mu}\vec{\pi}\right)  \cdot
\vec{\pi}\nonumber\\
&  -\frac{1}{4}(h_{2}^{\star}-2h_{3}^{\star})w_{a_{1}}w_{K_{1}}Z_{\pi}%
Z_{K}^{2}\,\eta_{N}^{E}\left(  \bar{K}^{0}\partial_{\mu}K^{0}\partial^{\mu}%
\pi^{0}-\sqrt{2}\bar{K}^{0}\partial_{\mu}K^{+}\partial^{\mu}\pi^{-} \right.
\nonumber\\
&  \left.  - K^{-}\partial_{\mu}K^{+}\partial^{\mu}\pi^{0}-\sqrt{2}%
K^{-}\partial_{\mu}K^{0}\partial^{\mu}\pi^{+}\text{ + h.c.}\right) \nonumber\\
&  -\frac{1}{2}h_{2}^{\star}w_{K_{1}}^{2}Z_{\pi}Z_{K}^{2}\,\eta_{N}^{E}\left(
\pi^{0}\partial_{\mu}\bar{K}^{0}\partial^{\mu}K^{0}-\pi^{0}\partial_{\mu}%
K^{-}\partial^{\mu}K^{+}-\sqrt{2}\pi^{-}\partial_{\mu}K^{+}\partial^{\mu}%
\bar{K}^{0}\text{ + h.c.}\right)  \;.
\end{align}

\subsection{Lagrangian for $\eta_{S}^{E}$}

The Lagrangian reads:%

\begin{align}
\mathcal{L}_{\eta_{S}^{E}}  &  =\,-\frac{i}{\sqrt{2}}h_{3}^{\star}w_{K_{1}%
}Z_{K}\phi_{N}\,\eta_{S}^{E}\left(  \partial_{\mu}\bar{K}^{0}K^{\star\mu
0}+\partial_{\mu}K^{-}K^{\star\mu+}\text{ + h.c.}\right)  +(h_{2}^{\star}-h_{3}^{\star
})w_{a_{1}}^{2}Z_{\pi}^{3}\,\eta_{N}^{E}\left(  \partial_{\mu}\eta_{N}%
\partial^{\mu}\vec{\pi}\right)  \cdot\vec{\pi}\nonumber\\
&  -\frac{1}{2\sqrt{2}}h_{2}^{\star}w_{a_{1}}w_{K_{1}}Z_{\pi}Z_{K}^{2}%
\,\eta_{S}^{E}\left(  \bar{K}^{0}\partial_{\mu}K^{0}\partial^{\mu}\pi
^{0}-\sqrt{2}\bar{K}^{0}\partial_{\mu}K^{+}\partial^{\mu}\pi^{-}\right.
\nonumber\\
&  \left.  -K^{-}\partial_{\mu}K^{+}\partial^{\mu}\pi^{0}-\sqrt{2}%
K^{-}\partial_{\mu}K^{0}\partial^{\mu}\pi^{+}\text{ + h.c.}\right) \nonumber\\
&  +\frac{1}{\sqrt{2}}h_{3}^{\star}w_{K_{1}}^{2}Z_{\pi}Z_{K}^{2}\,\eta_{S}%
^{E}\left(  \pi^{0}\partial_{\mu}\bar{K}^{0}\partial^{\mu}K^{0}+\pi
^{0}\partial_{\mu}K^{-}\partial^{\mu}K^{+}+\sqrt{2}\pi^{-}\partial_{\mu}%
K^{+}\partial^{\mu}\bar{K}^{0}\text{ + h.c.}\right)  \;.
\end{align}
$\,$
\\
Note: the decay $\eta_{S}^{E}\rightarrow\eta_{S}\pi\pi$ ($\sim\kappa_{1}$,
$h_{1}^{\star}$) is large-$N_{c}$ suppressed.

\subsection{Lagrangian for $\pi^{E}$}

The Lagrangian reads (only $\pi^{0E}$ included; decays of $\pi^{\pm E}$ follow
from isospin symmetry):%

\begin{align}
\mathcal{L}_{\pi^{E}}  &  =-\,ih_{3}^{\star}w_{a_{1}}Z_{\pi}\phi_{N}\,\pi
^{0E}\left(  \rho_{\mu}^{-}\partial^{\mu}\pi^{+}-\rho_{\mu}^{+}\partial^{\mu
}\pi^{-}\right) \nonumber\\
&  +\frac{1}{4}\left(  h_{2}^{\star}-2h_{3}^{\star}\right)  w_{a_{1}}w_{K_{1}%
}Z_{\pi}Z_{K}^{2}\,\pi^{0E}\partial_{\mu}\pi^{0}\left(  \bar{K}^{0}%
\partial^{\mu}K^{0}+K^{-}\partial^{\mu}K^{+}\text{ + h.c.}\right) \nonumber\\
&  +\frac{1}{2}h_{2}^{\star}w_{K_{1}}^{2}Z_{\pi}Z_{K}^{2}\,\pi^{0E}\pi
^{0}\left(  \partial_{\mu}\bar{K}^{0}\partial^{\mu}K^{0}+\partial_{\mu}%
K^{-}\partial^{\mu}K^{+}\right) \nonumber\\
&  -\frac{1}{2\sqrt{2}}\left(  h_{2}^{\star}+2h_{3}^{\star}\right)  w_{a_{1}%
}w_{K_{1}}Z_{\pi}Z_{K}^{2}\,\pi^{0E}\left[  \partial_{\mu}\pi^{-}\left(
\bar{K}^{0}\partial^{\mu}K^{+}-K^{+}\partial^{\mu}\bar{K}^{0}\right)  \text{ +
h.c.}\right] \nonumber\\
&  +\frac{1}{2}(h_{2}^{\star}+h_{3}^{\star})w_{a_{1}}^{2}Z_{\pi}^{3}\,\pi
^{0E}\pi^{0}(\partial_{\mu}\vec{\pi})^{2}-h_{3}^{\star}w_{a_{1}}^{2}Z_{\pi
}^{3}\,\pi^{0E}\partial_{\mu}\pi^{0}\left(  \vec{\pi}\cdot\partial^{\mu}%
\vec{\pi}\right)  \;.
\end{align}

\subsection{Lagrangian for $K^{E}$}

The Lagrangian reads (only $K^{0E}$ included; decays of other $K^{E}$
components follow from isospin symmetry):%

\begin{align}
\mathcal{L}_{K^{E}}  &  =-\frac{i}{4}\left[  h_{2}^{\star}\left(  \phi
_{N}
-
\sqrt{2}\phi_{S}\right) + 2h_{3}^{\star}\phi_{N}\right]  w_{K_{1}}%
Z_{K}\,K^{0E}\left(  \omega_{N\mu}\partial^{\mu}\bar{K}^{0}-\rho_{\mu}%
^{0}\partial^{\mu}\bar{K}^{0}+\sqrt{2}\rho_{\mu}^{+}\partial^{\mu}K^{-}\right)
\nonumber\\
&  -\frac{i}{4}\left[  h_{2}^{\star}\left(  \phi_{N}-\sqrt{2}\phi_{S}\right)
-2h_{3}^{\star}\phi_{N}\right]  w_{a_{1}}Z_{\pi}\,K^{0E}\left(  \bar{K}_{\mu
}^{\star0}\partial^{\mu}\eta_{N}-\bar{K}_{\mu}^{\star0}\partial^{\mu}\pi
^{0}+\sqrt{2}K_{\mu}^{\star-}\partial^{\mu}\pi^{+}\right) \nonumber\\
&  -\frac{i}{2\sqrt{2}}\left[  h_{2}^{\star}\left(  \phi_{N}-\sqrt{2}\phi
_{S}\right)  +2\sqrt{2}h_{3}^{\star}\phi_{S}\right]  w_{f_{1S}}Z_{\eta_{S}%
}\,K^{0E}\bar{K}_{\mu}^{\star0}\partial^{\mu}\eta_{S}\nonumber\\
&  -\frac{1}{2}h_{2}^{\star}w_{a_{1}}^{2}Z_{\pi}^{2}Z_{K}\,K^{0E}\left(
\bar{K}^{0}\partial_{\mu}\eta_{N}\partial^{\mu}\pi^{0}-\sqrt{2}K^{-}%
\partial_{\mu}\eta_{N}\partial^{\mu}\pi^{+}\right) \nonumber\\
&  -\frac{1}{4}(h_{2}^{\star}-2h_{3}^{\star})w_{a_{1}}w_{K_{1}}Z_{\pi}%
^{2}Z_{K}\,K^{0E}\left(  \pi^{0}\partial_{\mu}\eta_{N}\partial^{\mu}\bar
{K}^{0}-\sqrt{2}\pi^{+}\partial_{\mu}\eta_{N}\partial^{\mu}K^{-}+\eta
_{N}\partial_{\mu}\pi^{0}\partial^{\mu}\bar{K}^{0}-\sqrt{2}\eta_{N}%
\partial_{\mu}\pi^{+}\partial^{\mu}K^{-}\right) \nonumber\\
&  +\frac{1}{\sqrt{2}}h_{3}^{\star}w_{a_{1}}w_{f_{1S}}Z_{\pi}Z_{K}Z_{\eta_{S}%
}\,K^{0E}\left(  \bar{K}^{0}\partial_{\mu}\eta_{S}\partial^{\mu}\pi^{0}%
-\sqrt{2}K^{-}\partial_{\mu}\eta_{S}\partial^{\mu}\pi^{+}\right) \nonumber\\
&  -\frac{1}{2\sqrt{2}}h_{2}^{\star}w_{K_{1}}w_{f_{1S}}Z_{\pi}Z_{K}Z_{\eta
_{S}}\,K^{0E}\left(  \pi^{0}\partial_{\mu}\eta_{S}\partial^{\mu}\bar{K}%
^{0}-\sqrt{2}\pi^{+}\partial_{\mu}\eta_{S}\partial^{\mu}K^{-}\right)
\nonumber\\
&  -\frac{1}{2\sqrt{2}}h_{2}^{\star}w_{a_{1}}w_{K_{1}}Z_{\pi}Z_{K}Z_{\eta_{S}%
}\,K^{0E}\left(  \eta_{S}\partial_{\mu}\pi^{0}\partial^{\mu}\bar{K}^{0}%
-\sqrt{2}\eta_{S}\partial_{\mu}\pi^{+}\partial^{\mu}K^{-}\right) \nonumber\\
&  +\frac{1}{4}h_{2}^{\star}w_{a_{1}}^{2}Z_{\pi}^{2}Z_{K}\,K^{0E}\bar{K}%
^{0}(\partial_{\mu}\vec{\pi})^{2}+\frac{1}{4}(h_{2}^{\star}-2h_{3}^{\star
})w_{a_{1}}w_{K_{1}}Z_{\pi}^{2}Z_{K}\,K^{0E}\pi^{0}\partial_{\mu}\bar{K}%
^{0}\partial^{\mu}\pi^{0}\nonumber\\
&  -h_{3}^{\star}w_{a_{1}}w_{K_{1}}Z_{\pi}^{2}Z_{K}\,K^{0E}\pi^{-}%
\partial_{\mu}\bar{K}^{0}\partial^{\mu}\pi^{+}+\frac{1}{2}h_{2}^{\star
}w_{a_{1}}w_{K_{1}}Z_{\pi}^{2}Z_{K}\,K^{0E}\pi^{+}\partial_{\mu}\bar{K}%
^{0}\partial^{\mu}\pi^{-}\nonumber\\
&  +\frac{1}{2\sqrt{2}}(h_{2}^{\star}+2h_{3}^{\star})w_{a_{1}}w_{K_{1}}Z_{\pi
}^{2}Z_{K}\,K^{0E}\left(  \pi^{+}\partial_{\mu}K^{-}\partial^{\mu}\pi^{0}%
-\pi^{0}\partial_{\mu}K^{-}\partial^{\mu}\pi^{+}\right)  \;.
\end{align}

\end{document}